\documentclass[aps, aps,amsmath,amssymb,amsfonts, superscriptaddress,twocolumn, footinbib]{revtex4-2}
\usepackage[german,american,english]{babel}
\usepackage{graphicx}% Include figure files
\usepackage{graphics}% Include figure files
\usepackage{dcolumn}% Align table columns on decimal point
\usepackage{multirow}
\usepackage{bm}% bold math
\usepackage{latexsym}
\usepackage{amssymb}
\usepackage{amsmath}
\usepackage{amsfonts}
\usepackage{layout}
\usepackage{verbatim}
\usepackage{epsfig}
\usepackage{graphicx}
\usepackage{amsbsy}
\usepackage{xcolor}
\usepackage[caption=false]{subfig}
\newcommand{\bea}{\begin{eqnarray*}}
	\newcommand{\eea}{\end{eqnarray*}}
\newcommand{\bne}{\begin{equation*}}
\newcommand{\ede}{\end{equation*}}

\newcommand{\bnen}{\begin{equation}}
\newcommand{\eden}{\end{equation}}
\newcommand{\bean}{\begin{eqnarray}}
\newcommand{\eean}{\end{eqnarray}}
\newcommand{\bsen}{\begin{subequations}}
	\newcommand{\esen}{\end{subequations}}
\newcommand{\ba}{\arraycolsep 0.3ex \begin{array}{rl}}
\newcommand{\ea}{\end{array}}
\newcommand{\bna}{\begin{array}}
	\newcommand{\eda}{\end{array}}
\newcommand{\bnm}{\begin{enumerate}}
	\newcommand{\edm}{\end{enumerate}}

\newcommand {\ket} [1] {| #1 \rangle}

\def\pz{{\partial}}

\def\Hj{{\hat \jmath}}

\def\Hv{{\hat v}}

\def\HBr{\hat{\bm r}}

\def\HL{{\hat L}}

\def\Bk{{\bm k}}

\def\BE{{\bm E}}
\def\HH{{\hat H}}
\def\CR{{\mathcal R}}
\def\BCR{{\bm\CR}}
\def\Hrho{{\hat\rho}}

\def\eps{\epsilon}
\def\ve{{\varepsilon}}

\def\frac#1#2{{\textstyle{#1 \over #2}}}

\def\Der#1#2{{D #1\over D #2}}

\def\nd{^{\vphantom{\dagger}}}
\def\ns{^{\vphantom{*}}}

\def\half{\frac{1}{2}}

\def\hh{\hskip 0.1em}

\def\ket#1{{\big| \hh #1\hh \big\rangle}}

\begin{document}
	
\title{Giant orbital Hall effect due to the bulk states of 3D topological insulators}

\author{James H. Cullen}
\affiliation{School of Physics, The University of New South Wales, Sydney 2052, Australia}
\author{Hong Liu}
\affiliation{School of Physics, The University of New South Wales, Sydney 2052, Australia}
\author{Dimitrie Culcer}
\affiliation{School of Physics, The University of New South Wales, Sydney 2052, Australia}

\begin{abstract}
The highly efficient torques generated by 3D topological insulators make them a favourable platform for faster and more efficient magnetic memory devices. 
Recently, research into harnessing orbital angular momentum in orbital torques has received significant attention. Here we study the orbital Hall effect in topological insulators. We find that the bulk states give rise to a sizeable orbital Hall effect that is up to 3 orders of magnitude larger than the spin Hall effect in topological insulators. This is partially because the orbital angular momentum that each conduction electron carries is up to an order of magnitude larger than the $\hbar/2$ carried by its spin. Our results imply that the large torques measured in topological insulator/ferromagnet devices can be further enhanced through careful engineering of the heterostructure to optimise orbital-to-spin conversion. 
\end{abstract}

\date{\today}
\maketitle
\section{Introduction}

Recent years have seen a dramatic surge of interest in orbitronics \cite{Orbitronics-PRL-2005-Shoucheng, Orbitronics-in-action, Rhonald-Rev, cysne2025orbitronics}, whose focus is harnessing Bloch electrons' orbital angular momentum (OAM)\cite{Yafet-1963, Vanderbilt_2018} similarly to the way spintronics uses electron spin \cite{zutic_spintronicsrev, hirohata2020review, Roadmap-SOT-Review}. Orbitronics is primarily concerned with the generation of non-equilibrium orbital angular momentum densities and currents \cite{Exp-OHE-Ti-Nat-2023-Hyun-Woo,Hong-OHE-PRL,OHE-Binghai,OHE-PRB-2022-Manchon,OHE-metal-PRM-2022-Oppeneer,OHE-BiTMD-PRL-2021-Tatiana,RS-OHE-disorder,ISOHE-PRL-2018-Hyun-Woo,IOHE-Metal-PRB-2018-Hyun-Woo,OHE-Hetero-PRR-2022-Pietro,OHE-Weak-SOC-npj, PhysRevLett.131.156702, PhysRevB.106.184406,PhysRevLett.131.156703,10.1063/5.0106988,Exp-OEE-PRL-2022-Jinbo,Titov_EdgeOM}, a significant technological motivation being the electrical manipulation of magnetic degrees of freedom \cite{OT-FM-PRB-2021-YoshiChika,OT-OEE-NatComm-2018-Haibo,OT-PRR-2020-Hyun-Woo,OT-NatComm-2021-Kyung-Jin, Exp-OT-PRR-2020,Exp-OT-CommP-2021-Byong-Guk, LS-conversion-CP-2021-Byong-Guk,OOS-Cvert-2020-PRL-Mathias, PhysRevResearch.4.033037, OHE-OT-large, L-S-OT-2023}, with an emphasis on weakly spin-orbit coupled materials \cite{Exp-graphene-OHE-arXiv-2022, OAM-Exp-Tobias,OAM-PRL-2020-Reinert, OEE-NatComm-2019-Peter, Tangping, Inverse-OHE-weak-SOC,PhysRevResearch.6.013208}. The efficient control of magnetisation dynamics has potential applications in magnetic devices such as magnetic random-access memory (MRAM)\cite{Roadmap-SOT-Review, Ramaswamy2018, CI-SOT-RMP-2019-Manchon}, logic-in memory\cite{fan2017memory, wang2022all}, and neuromorphic computing devices \cite{marrows2024neuromorphic, grollier2020neuromorphic}.

Topological insulators (TI) are prime candidates for building magnetic torque devices\cite{SOT-TM-Rev}. Topological insulators have strong spin-orbit coupling and topologically protected chiral surface states that can produce a sizeable Rashba Edelstein effect\cite{Chang2015, PhysRevB.97.134402-Manchon}. Room-temperature magnetisation switching has been demonstrated in a number of TI/FM devices\cite{TI-FM-Switch,PhysRevLett.119.077702,TI-FM-Switch-NM, wang2023room} and recently the field-free operation of a TI MRAM device has also been demonstrated\cite{cui2023low}. Topological insulator spin torques are generally attributed to three mechanisms; the REE in the surface states, the spin Hall effect (SHE) in the bulk states and the spin-transfer torque (STT)\cite{kurebayashi2019}. Determining the dominant mechanisms in TI spin torques has historically been quite difficult\cite{SOT-TI-bulk-NL, TI-FM-Switch}. Recent calculations have shown the size of the spin Hall effect to be negligible\cite{Hong-PSHE-PRB, ma2024spin}, whereas the spin-transfer torque in the bulk states is potentially of a similar magnitude to the REE\cite{James-SOT}. Topological insulators have strong spin-orbit coupling, and should be expected to host both orbital Hall and Edelstein effects. The strong circular dichroism recorded in Bi$_2$Se$_3$ surface states indicates that they possess substantial chiral OAM\cite{PhysRevLett.108.046805}. Hence, the existence of a large orbital contribution to the torque in TI/FM systems should not be dismissed. There will be an orbital Edelstein effect (OEE) at the surface of the TI due to the topological surface states\cite{osumi2021kinetic}.%discuss in more detail?
There will also be an orbital Hall effect (OHE) due to bulk states, to date the OHE has not been calculated in topological insulators. The OHE refers to the generation of transverse orbital currents by an electric field. In this paper, we study the orbital degree of freedom of TI surface states and calculate the orbital Hall current in the TI bulk states.

In this work, we calculate the orbital Hall effect in two topological insulators Bi$_2$Se$_3$ and Sb$_2$Te$_3$, using the model Hamiltonian derived in Ref.\cite{TI-bulk-Parameters}, and compare its magnitude with the spin Hall effect in these materials. We also calculate the orbital angular momentum carried by the bulk states. We find that the orbital Hall effect in these materials is up to 3 orders of magnitude larger than the spin Hall effect. Showing that even in topological materials with strong spin orbit coupling we find that orbital effects overwhelm spin effects. We calculate the OHE using both the conventional and full definitions of the orbital current\cite{liu2024quantumcorrectionorbitalhall}, we find the orbital current to be large regardless of the definition. This difference in magnitude is partially explained by the fact that the OAM carried by each bulk electrom can be as large as $\sim$10$\hbar$. Lastly, we discuss the orbital-to-spin conversion and the orbital torque generated by the OHE using a phenomenological model\cite{CIAM-PRR-2020-Yuriy, OT-PRR-2020-Hyun-Woo}.

Our results lead to three important conclusions: (i) For any reasonable orbital-to-spin conversion efficiency (larger than $\sim$0.1$\%$) the orbital Hall torque will dominate the spin Hall torque; % I am not sure I believe in orbital-to-spin conversion. 
(ii) It is too early to tell if the orbital torque dominates the torque exerted on the magnetisation in the ferromagnets -- the OHE may give a sizeable contribution to the torque that could compete with the contribution from the REE in the surface states, but a quantitative comparison is challenging at the moment; (iii) However, regardless of this, future attempts at building efficient TI/FM torque devices should try to harness this large orbital current. Hence, optimising the orbital-to-spin conversion in TI spin torque devices is crucial for more efficient magnetisation control. This optimisation will likely require advanced interface-engineering techniques as well as specific choices of ferromagnets. Interface engineering has already been shown to be able to improve the spin torque efficiency in a TI spin torque device \cite{ojha2023spin}, as well has for orbital torque devices \cite{lyalin2024interface}. Furthermore, it is well known that choosing ferromagnets with strong spin-orbit coupling can greatly enhance the orbital torque\cite{OHE-OT-large,PhysRevResearch.5.023054}.

\begin{figure}[t]
\begin{center}
\includegraphics[trim=0cm 0cm 0cm 0cm, clip, width=0.8\columnwidth]{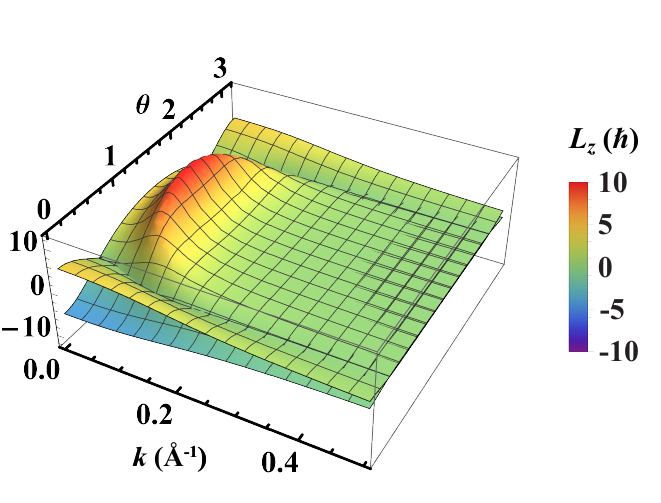}
\caption{\label{TI-bulk-Lz} The $L_z$ dependence on the ${\bm k}$-wavevector in equilibrium for the TI bulk states in Bi$_2$Se$_3$. (Bi$_2$Se$_3$ parameters from Ref.~\cite{TI-bulk-Parameters}).}
\end{center}
\end{figure}

\section{Results}
\begin{figure}[t]
\begin{center}
\includegraphics[trim=0cm 0cm 0cm 0cm, clip, width=0.8\columnwidth]{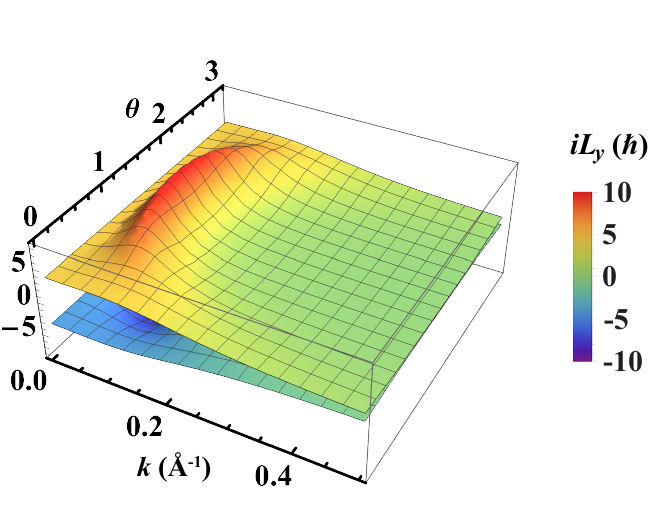}
\caption{\label{TI-bulk-Ly} The $i L_y$ dependence on the ${\bm k}$-wavevector in equilibrium for the TI bulk states in Bi$_2$Se$_3$. (Bi$_2$Se$_3$ parameters from Ref.~\cite{TI-bulk-Parameters}).}
\end{center}
\end{figure}

We applied our orbital current theory to the $4\times4$ TI bulk Hamiltonian in Ref.~\cite{TI-bulk-Parameters}. We calculated the orbital Hall conductivity $\sigma^y_{zx}$, where $\langle\Hj^y_{zx}\rangle=E_x\sigma^y_{zx}$, the orbital current is along $\hat{\bm z}$-direction carrying OAM aligned along $\hat{\bm y}$-direction with an applied electric field ${\bm E}$ along the $\hat{\bm x}$-direction. We also calculated the OAM in equilibrium for TI bulk Hamiltonian, this is shown in Fig.~\ref{TI-bulk-Lz} and Fig.~\ref{TI-bulk-Ly}, the system is isotropic in the $k_x$-$k_y$ plane so we plot the OAM vs $k$ and the polar angle $\theta$. The expression for the OAM of Bloch electrons contains the Berry curvature\cite{Theroy-OM-PRL-2007-Qian}, which tends to be large in strongly spin-orbit coupled materials. Accordingly, we find the magnitude of the OAM is much larger than that of spin angular momentum (SAM) $(1/2) \hbar$.

The decomposition of the OHE for the TI bulk Hamiltonian follows the notation in Ref.~\cite{liu2024quantumcorrectionorbitalhall}, the orbital current is split into two contributions the conventional term $j_{\text{conv}}$ and the quantum correction $\Delta j$.
The quantum correction $\Delta j$ can be split into three contributions $\Delta j_{1,2,3}$ \cite{liu2024quantumcorrectionorbitalhall}. 
The first contribution $\Delta j_1$ can be related to the dipole generated by the applied electric field displacing electrons away from their equilibrium center of mass. This dipole rotates, generating an OAM, and the OAM is then convected generating an orbital current. This mechanism can also be used to describe the conventional contribution. $\Delta j_2$ arises due to the interband matrix elements of the OAM operator. While these matrix elements do not contribute to the expectation value of the OAM in equilibrium, they do contribute to the orbital current. The last contribution to quantum correction $\Delta j_3$ arises due to the non-commutativity of the position and velocity operators. The orbital current does not require a charge current, thus it can be nonzero in the insulating state, an understanding reinforced by the fact that it is related to dipolar motion. Note that similar considerations apply to a spin current, which can also be nonzero in the gap\cite{ma2024spin}.

\begin{figure}[t]
\begin{center}
\includegraphics[trim=0cm 0cm 0cm 0cm, clip, width=\columnwidth]{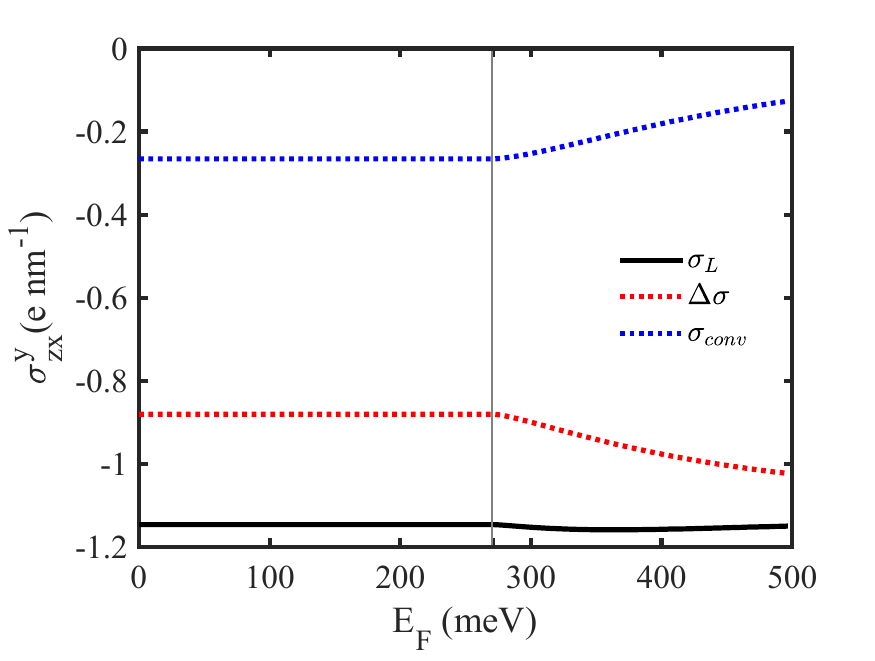}
\caption{\label{TI-bulk-zx-OHE1}  The orbital Hall conductivity vs $E_F$ for the TI bulk states in Bi$_2$Se$_3$. The contribution from the conventional term $\sigma_{conv}$, the contribution from the quantum correction $\Delta \sigma$ and the total $\sigma_L$ are plotted separately. The conduction band bottom is indicated by the vertical grey line. (Bi$_2$Se$_3$ parameters from Ref.~\cite{TI-bulk-Parameters}).}
\end{center}
\end{figure}

%Can we reference an equation we haven't shown yet?
In Fig.~\ref{TI-bulk-zx-OHE1}, we plot the conventional OHE conductivity and the quantum correction to OHE conductivity, we find the quantum correction to dominate the conventional term, this is consistent with our results in Ref.~\cite{liu2024quantumcorrectionorbitalhall} for the CuMnAs model. We find that in TIs $\Delta j_2$ is the dominant contribution to the orbital current. We also find the magnitude of $\Delta j_1$ to be significant, it is similar in size to $j_\text{conv}$ shown in Fig. \ref{TI-bulk-zx-OHE1}. This is unsurprising as they both originate from the diagonal and off-diagonal parts of the first term in our expression for the orbital current. We find the last contribution to the orbital current $\Delta j_3$ to be negligible. As shown in the figure we find the OHE to be non zero in the bulk band gap, once the Fermi energy is in the conduction band ($E_F>270$ meV) the orbital Hall conductivity will start to decrease. For comparison, we also plot the spin Hall current with both proper and conventional definitions. As is shown, the magnitude of the spin Hall conductivity is 2-3 orders of magnitude smaller than the orbital Hall conductivity, regardless of definition.

Non-equilibrium effects can be classified as intrinsic and extrinsic depending on whether they originate in the band structure or disorder. Terms that are zeroth order in the disorder strength are known to play crucial roles in Hall effects\cite{PhysRevLett.96.056602, PhysRevLett.104.186403, PhysRevLett.112.066601, PhysRevB.108.245418, Hong-OHE-PRL}. In the density matrix formalism, such corrections are incorporated into an \textit{anomalous driving term} \cite{Interband-Coherence-PRB-2017-Dimi} which results in a correction to $\rho^{mn}_{E{\bm k}}$ in Eq.~\ref{rhoE}. Here we estimate the extrinsic contribution to the full orbital current in the relaxation-time approximation, where the inverse relaxation time serves as a measure of the disorder strength. The extrinsic OHE will be zeroth-order in the relaxation time, that is, the same order as the intrinsic OHE \cite{Hong-OHE-PRL}. The relaxation time approximation does not take into account the electrical field corrected scattering integral \cite{JE-PRR-Rhonald-2022}, which is extremely laborious. Our results are shown in Fig.~\ref{TI-bulk-zx-OHE3}. We find that the extrinsic contribution is nearly same magnitude as the conventional OHE for the 3D TI bulk model and is hence much smaller than the intrinsic contribution $j_L$. Furthermore, we find the extrinsic contribution to have the same sign as the intrinsic contribution enhancing the total orbital current.

%% https://journals.aps.org/prresearch/abstract/10.1103/PhysRevResearch.2.033401
%%https://journals.aps.org/prresearch/abstract/10.1103/PhysRevResearch.2.013177

\section{Discussion}
We have calculated the orbital angular momentum and orbital Hall effect in the bulk states of topological insulators. We find that the magnitude of the orbital Hall effect dominates the spin Hall effect in both of the materials we studied Bi$_2$Se$_3$ and Sb$_2$Te$_3$. We show that one of the causes of this difference in magnitude is that while the size of the angular momentum carried by the spin of each bulk electron is fixed at $\hbar/2$ the orbital angular momentum is not, we find that it can be as large as $\sim$10$\hbar$. It has previously been shown that the orbital Hall conductivity can overwhelm the spin Hall conductivity in many materials\cite{ISHE-IOHE-PRB-2008-Inoue,OHE-PRL-2009-Inoue,IOHE-Metal-PRB-2018-Hyun-Woo,canonico2020two}, here we have shown that even in topological materials with strong spin-orbit coupling the orbital Hall conductivity dominates. Using the conventional definition, the orbital Hall current is 2 orders of magnitude larger than the spin Hall current, whereas using the complete definition the orbital Hall current is 3 orders of magnitude larger. Hence, as long as the orbital-to-spin conversion is greater than 1$\%$ we expect the orbital Hall effect to completely dominate the spin Hall effect in topological insulator/ferromagnetic torque devices.

\begin{figure}[t]
\begin{center}
\includegraphics[trim=0cm 0cm 0cm 0cm, clip, width=\columnwidth]{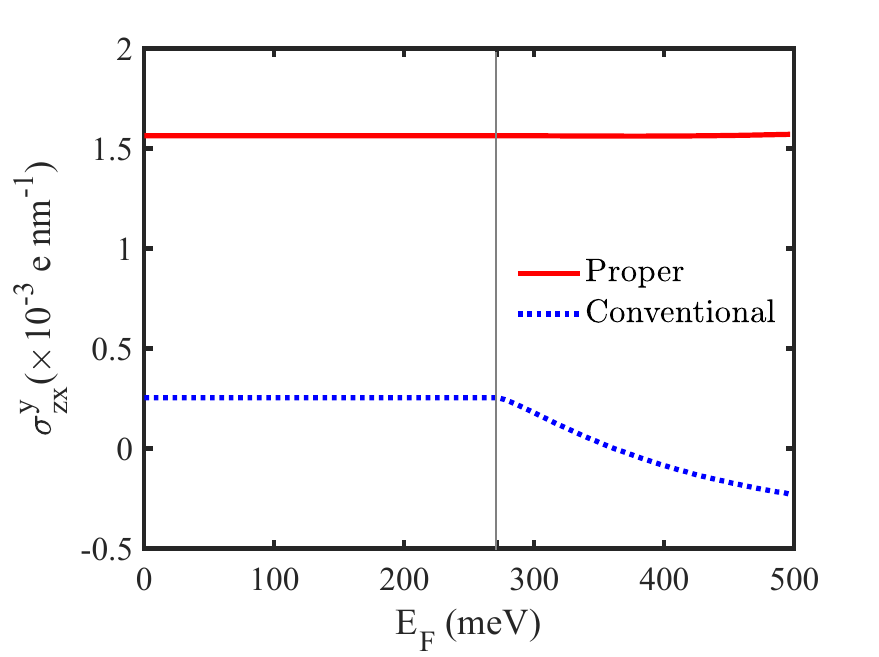}
\caption{\label{TI-bulk-zx-SHE} The intrinsic spin Hall conductivity $\sigma^{y}_{zx}$ vs $E_F$ for the TI bulk states in Bi$_2$Se$_3$, calculated using both proper and conventional definitions of the spin current. The conduction band bottom is indicated by the vertical grey line. (Bi$_2$Se$_3$ parameters from Ref.~\cite{TI-bulk-Parameters}).}
\end{center}
\end{figure}

The electrical manipulation of local magnetisation can be achieved via orbital and spin torques\cite{Roadmap-SOT-Review}. Spin torques occur due to the generation of a spin accumulation in a ferromagnetic material, if the spins are misaligned with the local magnetisation they will exert a torque on the magnetisation via the exchange interaction. Spin torques are usually induced via spin currents or spin densities generated at the interface of the ferromagnet (FM) with a non-magnetic material. Orbital torques refer to a similar phenomenon in which orbital densities and currents are generated in an adjacent non-magnetic material and a torque is induced on the magnetisation of the ferromagnet. If the size of the orbital/spin torque is large enough it can be used to manipulate magnetic textures or switch the magnetisation. However, the exact mechanisms through which the orbital torque is generated is still unclear as the OAM cannot directly interact with the local magnetisation -- it does not participate in the exchange interaction. The current understanding of magnetisation dynamics due to the orbital torque consists of three steps: The generation of an orbital current/density in the TI layer which is then injected into the FM layer. Secondly, the OAM is converted to spin by the spin-orbit coupling (SOC) in the FM layer. Finally, the spins in the FM layer exert a torque on the local magnetisation\cite{OT-PRR-2020-Hyun-Woo, CIAM-PRR-2020-Yuriy}. This mechanism has experimental evidence, by comparing the sign of the induced torque in different materials with the sign of the materials' spin-orbit coupling parameter\cite{Exp-OT-CommP-2021-Byong-Guk, OHE-Hetero-PRR-2022-Pietro}.

We find that the orbital conductivities in the materials we studied are of the order $10^5-10^6(\hbar/2e)\Omega^{-1}\text{m}^{-1}$ shown in Fig. \ref{OHE-combined}, which is significant as TI spin torque experiments find their spin conductivities to be of the order $10^4-10^6(\hbar/2e)\Omega^{-1}\text{m}^{-1}$\cite{PhysRevB.97.134402-Manchon,SOT-TM-Rev}. Hence, the large orbital Hall effect in the TI bulk states presents another avenue for the enhancement of the torque efficiency in TI devices. It is already known that the topological surface states generate a large spin torque via the Rashba-Edelstein effect\cite{Chang2015, PhysRevB.97.134402-Manchon}. This torque mechanism should exist in any TI device, whereas the orbital torque requires good orbital-to-spin conversion to be significant. We propose that TI/FM torques could be further enhanced by taking advantage of the large OHE, through interface engineering and careful choices of ferromagnet that would improve the orbital-to-spin conversion. Additionally, the FM will also need a spin-orbit coupling parameter with the correct sign such that the spins converted from the orbital current align with the spins generated from the REE.

\begin{figure}[t]
\begin{center}
\includegraphics[trim=0cm 0cm 0cm 0cm, clip, width=\columnwidth]{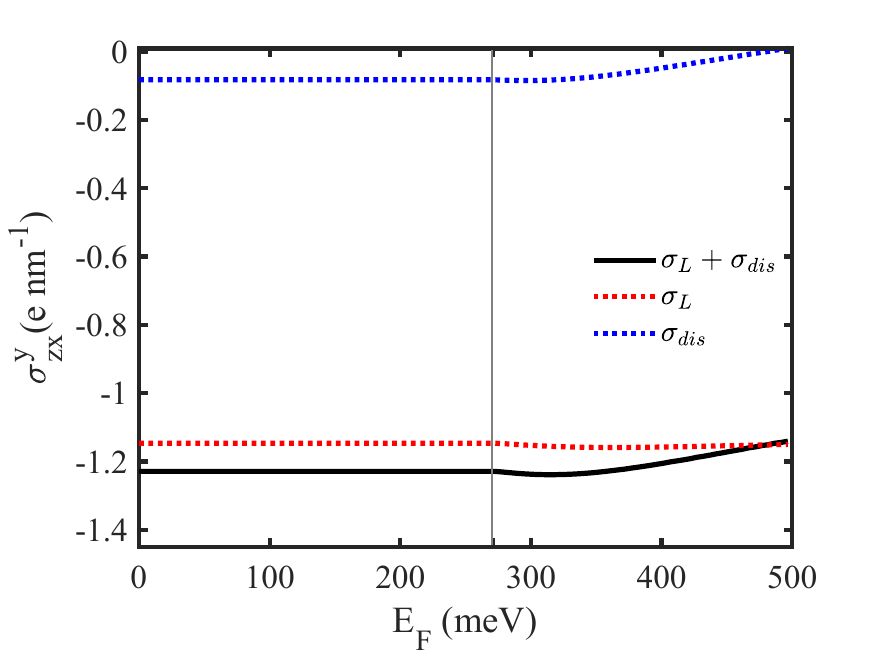}
\caption{\label{TI-bulk-zx-OHE3} The orbital Hall conductivity vs $E_F$ for the TI bulk states in Bi$_2$Se$_3$. Here the intrinsic $\sigma_L$, extrinsic $\sigma_{\rm dis}$, and total OHE have been plotted separately. The conduction band bottom is indicated by the vertical grey line. (Bi$_2$Se$_3$ parameters from Ref.~\cite{TI-bulk-Parameters}).}
\end{center}
\end{figure}

Spin conductivity due to the Rashba-Edelstein effect in Bi$_2$Se$_3$ has been shown to be on the order of $10^4-10^5 (\hbar/2e) \Omega^{-1}$/m\cite{PhysRevB.97.134402-Manchon}, this effect is believed to be the primary driver behind the large spin torques measured in TI/FM devices. However, it was recently shown that there will be a spin-transfer torque due to the bulk states that could potentially be of a similar magnitude to the REE\cite{James-SOT}. The spin Hall conductivity shown in Fig. 4 is of the order $10^3(\hbar/2e) \Omega^{-1}$/m. Furthermore, a recent calculation employing an ab-initio model showed that the intrinsic spin Hall conductivity to be $\sigma^y_{zx}=-2.2\times10^3(\hbar/2e) \Omega^{-1}$/m \cite{ma2024spin}. Hence, the intrinsic spin Hall effect is likely negligible in TI spin torques. The extrinsic spin Hall conductivity has also been shown to be of negligible magnitude\cite{cullen2023spin}. Hence, the magnitude of the orbital Hall effect is large in the context of the other known spin torque mechanisms in TI/FM devices. 

The role of the orbital Edelstein effect in TI/FM spin torques remains to be elucidated. In general, it is only possible for the OEE in the surface states to generate an orbital magnetisation along the out-of-plane direction as electrons are confined along this axis. The simplest model to describe the surface states at the interface in a TI/FM device would be a massive Dirac cone $H=\alpha(\boldsymbol{k}\times\boldsymbol{\sigma})_z-m\sigma_z$, our numerical estimates of the OEE ($\boldsymbol{L}\parallel\hat{z}$) in this system find it to be zero. However, the OEE has been calculated for a 2D topological insulator\cite{osumi2021kinetic} and was shown to be large. There has also been recent theoretical work showing that an in-plane OEE can be electrically induced in the bulk states of materials in heterostructures due to inversion asymmetry\cite{OEE-scalar-potential} or interface reflections\cite{Titov_EdgeOM}, although it is unclear how large this effect would be in a TI/FM device and whether the induced magnetisation could generate a spin torque.

\begin{figure}[t]
\begin{center}
\includegraphics[trim=0cm 0cm 0cm 0cm, clip, width=\columnwidth]{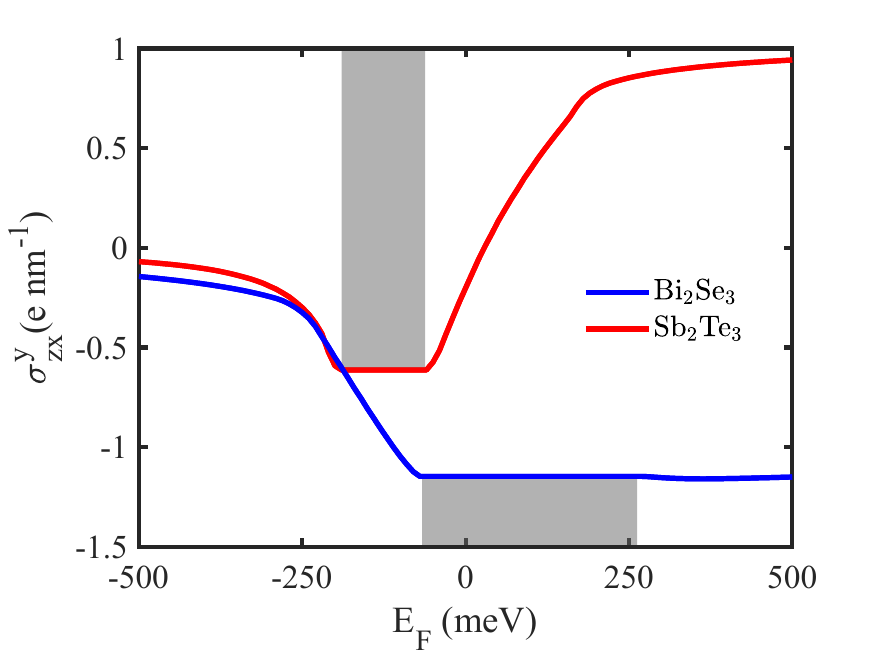}
\caption{\label{OHE-combined} Intrinsic orbital Hall conductivity for Bi$_2$Se$_3$ and Sb$_2$Te$_3$. The top of the valence band and bottom of the conduction band for each material are indicated the shaded areas. Parameters are from Ref.~\cite{TI-bulk-Parameters}.}
\end{center}
\end{figure}

The relative roles of the surface and bulk states in TI spin torques have historically been unclear. However, experimental results along with recent theoretical evidence imply that spin torques are largely dominated by surface state and interface effects. It has been demonstrated that when decreasing the Fermi energy via doping there is an increase in the torque efficiency up to a certain point before it starts to decrease again\cite{PhysRevLett.123.207205}, this behaviour indicates that the spin torque efficiency increases when the Fermi energy is in the gap and near the Dirac point. Furthermore, another experiment showed that the spin torque efficiency is greatly enhanced in thinner TI samples\cite{TI-FM-Switch}. Recently, the proper spin current has been calculated in the bulk of TI's and shown to be quite small\cite{Hong-PSHE-PRB, ma2024spin}. All of this evidence seems to imply that the spin torque contribution coming from the bulk is negligible. However, this analysis does not straightforwardly apply to the orbital torque as orbital-to-spin conversion is not considered. So, while the bulk contribution to the spin torque should be negligible, the same cannot be said for the orbital torque contribution from the bulk. In fact, some of the larger spin torques measured in TI/FM devices use the FM CoFeB\cite{dc2018room,wu2021magnetic,wu2019room}, a FM that has been shown to have a reasonable orbital-to-spin conversion\cite{OT-NatComm-2021-Kyung-Jin}.

The OHE will inject an orbital current into the FM layer, this then generates an orbital accumulation in the FM. The orbital accumulation in the FM layer is thought to generate a spin accumulation via spin-orbit coupling. The exact mechanism through which this occurs is unclear, it has been proposed that the SOC of the atomic orbitals is responsible\cite{OT-PRR-2020-Hyun-Woo}. The SOC in the FM layer can be written as $H^\text{FM}_\text{so}=\alpha^\text{FM}_\text{so} {\bm L}\cdot{\bm S}$, where $\alpha^\text{FM}_\text{so} $ is the SOC coupling coefficient, the spins generated will either be aligned parallel or anti-parallel to the orbital accumulation depending on the sign of the coefficient. The exchange coupling between the magnetisation ${\bm M}$ in the FM and spin is $H^\text{FM}_\text{xc}=J{\bm M}\cdot{\bm S}$, where $J$ is the exchange coupling, it is this exchange coupling that causes magnetisation dynamics. If the spins generated are misaligned with the magnetisation they will generate a torque on the magnetisation ${\bm T}= \tau {\bm M}\times \hat{\bm S}$. 

Additionally to mechanism mentioned above, there will also be a contribution to the orbital torque due to the orbital accumulation being converted to a spin accumulation via SOC within the TI layer, this spin accumulation will diffuse into the FM and also generate a torque. Lastly, the interface between the FM and NM is also known to be important for orbital-to-spin conversion in orbital torques\cite{OT-PRR-2020-Hyun-Woo}, however, the details of the physics at the interface is very much an open problem. A recent experiment showed evidence that the interface transparency to orbital currents is often greater than for spin currents \cite{lyalin2024interface}, the paper also demonstrated orbital torque enhancement via interface engineering.

Each bulk electron can carry a large amount of orbital angular momentum as shown in Figures \ref{TI-bulk-Lz} and \ref{TI-bulk-Ly}. Additionally, an applied electric field will further generate orbital angular momentum on the level of the wave packet. Unlike the angular momentum of the electron spin which is fixed at $\hbar/2$ the orbital angular momentum does not have this restriction. Not only does this partially explain why the OHE dominates the SHE in the TI bulk but it gives further motivation to the pursuit of orbitronics, as it shows that harnessing the OAM could present a more efficient way to build spintronic devices. In order to produce the spin densities and spin currents used in spin torques, spin-orbit coupling is normally required. It is known that spin-orbit coupling is also a source of OAM \cite{OEE-scalar-potential}. So, it is likely that in most devices there will be a combination of spin and orbital torques. Furthermore, generally the effective spin Hall conductivity obtained from spin torque measurements does not distinguish between orbital and spin contributions. Recent theories imply that the current-induced dynamics and spin transport in the presence of spin-orbit coupling originate in the orbital degrees of freedom\cite{PhysRevLett.102.016601, PhysRevLett.121.086602}. Furthermore, it has been shown that the Rashba-Edelstein effect (REE), in which a electrically induced spin polarisation is generated in a 2D system with Rashba spin-orbit coupling, is smaller than its orbital counterpart the Orbital-Edelstein effect (OEE)\cite{OAM-Rashba-PRL-2011-Changyoung,OEE-scalar-potential}. This, along with the large OHE we calculated in this work solidifies the need to pursue the ability to engineer devices with better orbital-to-spin conversion.

It should be mentioned that despite the calculated orbital Hall current being large, relating the orbital current directly to the torque is non-trivial. As mentioned previously, there is already the complication of orbital-to-spin conversion, a topic that we still only have a rudimentary understanding of. In addition, even relating the orbital current to the orbital accumulation is difficult, this is already a known problem with the spin accumulation and the spin Hall effect\cite{Tatara-PRB-2018,Tatara-PRB-Letter,Hong-PSHE-PRB}. As has been done in this work and in previous works on the spin Hall effect, the best way around this is phenomenological and qualitative descriptions of the physics involved. In principle the total angular momentum is the most relevant physical quantity. Nevertheless, at the moment, the prescription for calculating this quantity for delocalised Bloch electrons has not been developed, and this remains an open fundamental question in the field.

We would like to emphasise that our work solely focuses on the bulk states of topological insulators. The orbital current we calculate is a bulk state effect. 
The behaviour of the orbital current vs the Fermi energy observed here is similar to that of CuMnAs studied in Ref.\cite{liu2024quantumcorrectionorbitalhall}, which is not known to have topological surface states. The topological surface state contribution to the orbital current $\Hj^y_{zx}$ must be zero as these states can only generate orbital currents that flow in-plane $\parallel \hat{x},\hat{y}$, and, additionally, they can only generate an OAM $\boldsymbol{L}\parallel \hat{z}$. Lastly, the orbital current is not explicitly related to any topological quantities, such as the Berry curvature, which would indicate a potential surface state origin \textit{hidden} in the bulk calculation. Hence, any orbital current of the form calculated in this paper, $\boldsymbol{L}$ in-plane and flow out-of-plane, can only be due to the bulk states. 

%We have studied the orbital Hall effect in the bulk of topological insulators. We find that the magnitude of the orbital Hall current is 2-3 orders of magnitude larger than the spin Hall current. We believe that this discovery could open new avenues for the enhancement of the torque efficiency in topological insulator/ferromagnet devices.

\section{Methods}
In this section, we outline the method for calculating the orbital current and the model Hamiltonian used.
\subsection{Orbital angular momentum and orbital current operators}
The evaluation of the orbital current is nontrivial in periodic solids because the position operator is ill-defined in extended systems \cite{Hong-PSHE-PRB}. The standard approach adopted to circumvent this problem is to start from the equilibrium matrix elements of the OAM operator and combine those with a non-equilibrium distribution found using standard methods such as the Boltzmann equation or the Kubo formula. This approach is incomplete and neglects important terms\cite{liu2024quantumcorrectionorbitalhall}, we refer to this as the conventional orbital current. The complete expression can only be obtained via a full quantum mechanical evaluation of the non-equilibrium expectation value of the orbital current operator, including all the resulting matrix elements of the position operator. This introduces extra contributions to the orbital current coined as the quantum correction\cite{liu2024quantumcorrectionorbitalhall}. In this work, in order to facilitate a comprehensive comparison, we calculate the orbital current using both the conventional definition and the corrected definition that includes the quantum correction.

Our evaluation of the orbital current and OAM follows the calculation in Ref.~\cite{liu2024quantumcorrectionorbitalhall}. We define the OAM operator as the symmetrized combination $\hat{\bm L}=\frac{1}{2}(\hat{\bm r}\times\hat{\bm v} - \hat{\bm v}\times\hat{\bm r})$ and the orbital current operator as $\Hj^\alpha_\delta = \half\big\{\HL_\alpha,\Hv_\delta\big\}$, where $\hat{\bm v}$ is the velocity operator. The expectation values of these operators are evaluated by taking the trace with the density matrix. We work in the Hilbert space spanned by Bloch wave-functions $\ket{\Psi_{m{\bm k}} } = e^{i{\bm k}\cdot{\bm r}} \ket{u_{m{\bm k}}}$. To evaluate the full OHE we require the non-equilibrium correction to the density matrix in an electric field, for which we use the linear response theory following the approach of Refs.~\cite{Interband-Coherence-PRB-2017-Dimi, JE-PRR-Rhonald-2022}. The single-particle density operator obeys the quantum Liouville equation, $\pz\Hrho/\pz t + (i/\hbar)[\HH,\Hrho]=0$, where $\HH=\HH_0+\HH_E$. Here $\HH_0$ is the band Hamiltonian and $\HH_E=e\BE\cdot\HBr$ is the potential due to the external electrical field. At this stage, we focus on intrinsic effects and do not consider disorder scattering, which will be discussed in closing. In the crystal momentum representation the equilibrium density matrix has the diagonal form $\rho_{0\Bk}^{mn} = f_m\, \delta_{mn}$, where $f_m \equiv f(\ve_{m{\bm k}})$ is the Fermi-Dirac distribution for band $m$. In an electric field the density matrix can be written as $\hat{\rho} = \rho_0+\rho_E$, and, in linear response, it has been shown that \cite{Interband-Coherence-PRB-2017-Dimi} 
\begin{equation}\label{rhoE}
\rho^{mn}_{E{\bm k}} = {f(\ve_{m{\bm k}})-f(\ve_{n{\bm k}})\over
\ve_{m{\bm k}} -\ve_{n{\bm k}} }\> e{\bm E}\cdot \BCR^{mn}_\Bk\quad,
\end{equation}
where $\boldsymbol{\mathcal{R}}_{\bm k}^{mn}=\langle u_{n\bm k}|i\partial u_{m\bm k}/\partial \bm k\rangle$ is the Berry connection. Once $\rho^{mn}_{E{\bm k}}$ is found the expectation value of the orbital current can be written as
\begin{equation}\label{OC}
\begin{aligned}
    \langle\Hj^\alpha_\delta\rangle =& \frac{\eps\ns_{\alpha\beta\gamma}}{4}  \sum_{m,\Bk} \big\{\CR\nd_\beta, \rho\nd_{E\Bk}\big\}^{mm}\, \big\{v_\delta, v_\gamma\big\}^{mm} +\\
    &{2eE_\mu \left[\Der{\Xi^0_\beta}{k_\mu}\right]^{mn} +\big\{\hbar v\nd_\beta, \rho\nd_{E\Bk}\big\}^{mn} \over \ve_n-\ve_m} \{v_\gamma, v_\delta\}^{nm}\\
    &+i\frac{\eps\ns_{\alpha\beta\gamma}}{4} \sum_{m \ne n,\Bk}\Big[v_\gamma, \Der{v_\delta}{k_\beta} \Big]^{mn}_\Bk \rho^{nm}_{E\Bk} \,,
\end{aligned}
\end{equation}
where we have abbreviated $\big[\Xi^0_\beta\big]^{mn}=\half\CR^{mn}_\beta (f_m + f_n)$, and the covariant derivative $DO/Dk_j=\partial O/\partial k_j - i[\mathcal{R}_j, O]$. 
% and $m \ne n$ is understood in the second summation. 
The OAM polarization is taken to be along the $\alpha$-direction while the transport direction is denoted by $\delta$. The expression used to calculate the orbital current was derived in Ref.~\cite{liu2024quantumcorrectionorbitalhall} and shown to be gauge invariant. The expression in (\ref{OC}) contains the quantum correction to the orbital current $\Delta j$ that arises due to the inclusion of all matrix elements, intra-band and inter-band, of the position and velocity operators. The intra-band elements of the position operator require careful consideration as they are differential operators and a full evaluation often requires accounting for elements off-diagonal in the wave vector \cite{Hong-PSHE-PRB,Rhonald-Conservation-OMM}. The conventional part of the orbital current is contained in the first term of (\ref{OC}), but only contains the off-diagonal components of the velocity operators, while all other terms in (\ref{OC}) constitute the quantum correction. The second line of (\ref{OC}) contains $\Delta j_2$ and the third line contains $\Delta j_3$. The part of the first line of (\ref{OC}) containing band diagonal components of the velocity operator is $\Delta j_1$. 
\subsection{Model Hamiltonian}
We apply our theory to the $4\times4$ TI bulk Hamiltonian in Ref.~\cite{TI-bulk-Parameters}, $H_{0{\bm k}}=\epsilon_{\bm k}+H_\text{so}$ where $\epsilon_{\bm k}=C_0+C_1k^2_z+C_2k^2_{\parallel}$. The spin-orbit coupling Hamiltonian is
\begin{equation}
\ba
H_\text{so}=
\left(
\begin{array}{cccc}
-\mathcal{M} & 0  & \mathcal{B}k_z  & \mathcal{A}k_-\\
  0 & -\mathcal{M}  &  \mathcal{A}k_+ & -\mathcal{B}k_z\\
\mathcal{B}k_z  & \mathcal{A}k_-  &  \mathcal{M} & 0\\
\mathcal{A}k_+  &   -\mathcal{B}k_z & 0  & \mathcal{M}
\end{array}
\right).
\ea
\end{equation}
The Hamiltonian is in basis $\{\frac{1}{2},-\frac{1}{2},\frac{1}{2},-\frac{1}{2}\}$. $\mathcal{M} = M_0+M_1k^2_z+M_2k^2_\parallel,  \mathcal{A}=A_0+A_2k^2_\parallel, \mathcal{B}=B_0+B_2k^2_z$. The wave-vector ${\bm k}=(k\sin\theta\cos\phi, k\sin\theta\sin\phi, k\cos\theta)$ with $\theta$ the polar angle and $\phi$  azimuthal angle. 
The Hamiltonian is in basis $\{\frac{1}{2},-\frac{1}{2},\frac{1}{2},-\frac{1}{2}\}$. $\mathcal{M} = M_0+M_1k^2_z+M_2k^2_\parallel,  \mathcal{A}=A_0+A_2k^2_\parallel, \mathcal{B}=B_0+B_2k^2_z$. 

\subsection*{Data Availability}
The authors confirm that the data supporting the findings of this study are available within the article and its supplementary materials.

\subsection*{Acknowledgements}
This work is supported by the Australian Research Council Discovery Project DP2401062. JHC acknowledges support from an Australian Government Research Training Program (RTP) Scholarship.

\subsection*{Author Contributions}
All authors wrote the main manuscript text and worked on the theory for the orbital Hall effect. J.H.C. and H.L. generated and prepared the figures in the manuscript.

\subsection*{Competing interests}
The authors declare no competing interests.

\subsection*{Additional Information}
Supplementary Information is available for this paper.

%\bibliographystyle{apsrev4-2}
%\bibliography{ref-TI-OHE}

\begin{thebibliography}{93}%
	\makeatletter
	\providecommand \@ifxundefined [1]{%
		\@ifx{#1\undefined}
	}%
	\providecommand \@ifnum [1]{%
		\ifnum #1\expandafter \@firstoftwo
		\else \expandafter \@secondoftwo
		\fi
	}%
	\providecommand \@ifx [1]{%
		\ifx #1\expandafter \@firstoftwo
		\else \expandafter \@secondoftwo
		\fi
	}%
	\providecommand \natexlab [1]{#1}%
	\providecommand \enquote  [1]{``#1''}%
	\providecommand \bibnamefont  [1]{#1}%
	\providecommand \bibfnamefont [1]{#1}%
	\providecommand \citenamefont [1]{#1}%
	\providecommand \href@noop [0]{\@secondoftwo}%
	\providecommand \href [0]{\begingroup \@sanitize@url \@href}%
	\providecommand \@href[1]{\@@startlink{#1}\@@href}%
	\providecommand \@@href[1]{\endgroup#1\@@endlink}%
	\providecommand \@sanitize@url [0]{\catcode `\\12\catcode `\$12\catcode
		`\&12\catcode `\#12\catcode `\^12\catcode `\_12\catcode `\%12\relax}%
	\providecommand \@@startlink[1]{}%
	\providecommand \@@endlink[0]{}%
	\providecommand \url  [0]{\begingroup\@sanitize@url \@url }%
	\providecommand \@url [1]{\endgroup\@href {#1}{\urlprefix }}%
	\providecommand \urlprefix  [0]{URL }%
	\providecommand \Eprint [0]{\href }%
	\providecommand \doibase [0]{https://doi.org/}%
	\providecommand \selectlanguage [0]{\@gobble}%
	\providecommand \bibinfo  [0]{\@secondoftwo}%
	\providecommand \bibfield  [0]{\@secondoftwo}%
	\providecommand \translation [1]{[#1]}%
	\providecommand \BibitemOpen [0]{}%
	\providecommand \bibitemStop [0]{}%
	\providecommand \bibitemNoStop [0]{.\EOS\space}%
	\providecommand \EOS [0]{\spacefactor3000\relax}%
	\providecommand \BibitemShut  [1]{\csname bibitem#1\endcsname}%
	\let\auto@bib@innerbib\@empty
	%</preamble>
	\bibitem [{\citenamefont {Bernevig}\ \emph {et~al.}(2005)\citenamefont
		{Bernevig}, \citenamefont {Hughes},\ and\ \citenamefont
		{Zhang}}]{Orbitronics-PRL-2005-Shoucheng}%
	\BibitemOpen
	\bibfield  {author} {\bibinfo {author} {\bibfnamefont {B.~A.}\ \bibnamefont
			{Bernevig}}, \bibinfo {author} {\bibfnamefont {T.~L.}\ \bibnamefont
			{Hughes}},\ and\ \bibinfo {author} {\bibfnamefont {S.-C.}\ \bibnamefont
			{Zhang}},\ }\href {https://doi.org/10.1103/PhysRevLett.95.066601} {\bibfield
		{journal} {\bibinfo  {journal} {Phys. Rev. Lett.}\ }\textbf {\bibinfo
			{volume} {95}},\ \bibinfo {pages} {066601} (\bibinfo {year}
		{2005})}\BibitemShut {NoStop}%
	\bibitem [{\citenamefont {Das}(2023)}]{Orbitronics-in-action}%
	\BibitemOpen
	\bibfield  {author} {\bibinfo {author} {\bibfnamefont {D.}~\bibnamefont
			{Das}},\ }\href {https://doi.org/10.1038/s41567-023-02183-4} {\bibfield
		{journal} {\bibinfo  {journal} {Nature Physics}\ }\textbf {\bibinfo {volume}
			{19}},\ \bibinfo {pages} {1085} (\bibinfo {year} {2023})}\BibitemShut
	{NoStop}%
	\bibitem [{\citenamefont {Burgos~Atencia}\ \emph {et~al.}(2024)\citenamefont
		{Burgos~Atencia}, \citenamefont {Agarwal},\ and\ \citenamefont
		{Culcer}}]{Rhonald-Rev}%
	\BibitemOpen
	\bibfield  {author} {\bibinfo {author} {\bibfnamefont {R.}~\bibnamefont
			{Burgos~Atencia}}, \bibinfo {author} {\bibfnamefont {A.}~\bibnamefont
			{Agarwal}},\ and\ \bibinfo {author} {\bibfnamefont {D.}~\bibnamefont
			{Culcer}},\ }\href@noop {} {\bibfield  {journal} {\bibinfo  {journal}
			{Advances in Physics: X}\ }\textbf {\bibinfo {volume} {9}},\ \bibinfo {pages}
		{2371972} (\bibinfo {year} {2024})}\BibitemShut {NoStop}%
	\bibitem [{\citenamefont {Cysne}\ \emph {et~al.}(2025)\citenamefont {Cysne},
		\citenamefont {Canonico}, \citenamefont {Costa}, \citenamefont {Muniz},\ and\
		\citenamefont {Rappoport}}]{cysne2025orbitronics}%
	\BibitemOpen
	\bibfield  {author} {\bibinfo {author} {\bibfnamefont {T.~P.}\ \bibnamefont
			{Cysne}}, \bibinfo {author} {\bibfnamefont {L.~M.}\ \bibnamefont {Canonico}},
		\bibinfo {author} {\bibfnamefont {M.}~\bibnamefont {Costa}}, \bibinfo
		{author} {\bibfnamefont {R.}~\bibnamefont {Muniz}},\ and\ \bibinfo {author}
		{\bibfnamefont {T.~G.}\ \bibnamefont {Rappoport}},\ }\href@noop {} {\bibfield
		{journal} {\bibinfo  {journal} {arXiv preprint arXiv:2502.12339}\ }
		(\bibinfo {year} {2025})}\BibitemShut {NoStop}%
	\bibitem [{\citenamefont {Yafet}(1963)}]{Yafet-1963}%
	\BibitemOpen
	\bibfield  {author} {\bibinfo {author} {\bibfnamefont {Y.}~\bibnamefont
			{Yafet}},\ }\href@noop {} {\bibfield  {journal} {\bibinfo  {journal} {Solid
				State Physics Eds. Seitz and Turnbull}\ }\textbf {\bibinfo {volume} {14}},\
		\bibinfo {pages} {1} (\bibinfo {year} {1963})}\BibitemShut {NoStop}%
	\bibitem [{\citenamefont {Vanderbilt}(2018)}]{Vanderbilt_2018}%
	\BibitemOpen
	\bibfield  {author} {\bibinfo {author} {\bibfnamefont {D.}~\bibnamefont
			{Vanderbilt}},\ }\href@noop {} {\emph {\bibinfo {title} {Berry Phases in
				Electronic Structure Theory: Electric Polarization, Orbital Magnetization and
				Topological Insulators}}}\ (\bibinfo  {publisher} {Cambridge University
		Press},\ \bibinfo {year} {2018})\BibitemShut {NoStop}%
	\bibitem [{\citenamefont {{\v{Z}}uti{\'c}}\ \emph {et~al.}(2004)\citenamefont
		{{\v{Z}}uti{\'c}}, \citenamefont {Fabian},\ and\ \citenamefont
		{Sarma}}]{zutic_spintronicsrev}%
	\BibitemOpen
	\bibfield  {author} {\bibinfo {author} {\bibfnamefont {I.}~\bibnamefont
			{{\v{Z}}uti{\'c}}}, \bibinfo {author} {\bibfnamefont {J.}~\bibnamefont
			{Fabian}},\ and\ \bibinfo {author} {\bibfnamefont {S.~D.}\ \bibnamefont
			{Sarma}},\ }\href@noop {} {\bibfield  {journal} {\bibinfo  {journal} {Reviews
				of modern physics}\ }\textbf {\bibinfo {volume} {76}},\ \bibinfo {pages}
		{323} (\bibinfo {year} {2004})}\BibitemShut {NoStop}%
	\bibitem [{\citenamefont {Hirohata}\ \emph {et~al.}(2020)\citenamefont
		{Hirohata}, \citenamefont {Yamada}, \citenamefont {Nakatani}, \citenamefont
		{Prejbeanu}, \citenamefont {Di{\'e}ny}, \citenamefont {Pirro},\ and\
		\citenamefont {Hillebrands}}]{hirohata2020review}%
	\BibitemOpen
	\bibfield  {author} {\bibinfo {author} {\bibfnamefont {A.}~\bibnamefont
			{Hirohata}}, \bibinfo {author} {\bibfnamefont {K.}~\bibnamefont {Yamada}},
		\bibinfo {author} {\bibfnamefont {Y.}~\bibnamefont {Nakatani}}, \bibinfo
		{author} {\bibfnamefont {I.-L.}\ \bibnamefont {Prejbeanu}}, \bibinfo {author}
		{\bibfnamefont {B.}~\bibnamefont {Di{\'e}ny}}, \bibinfo {author}
		{\bibfnamefont {P.}~\bibnamefont {Pirro}},\ and\ \bibinfo {author}
		{\bibfnamefont {B.}~\bibnamefont {Hillebrands}},\ }\href@noop {} {\bibfield
		{journal} {\bibinfo  {journal} {Journal of Magnetism and Magnetic Materials}\
		}\textbf {\bibinfo {volume} {509}},\ \bibinfo {pages} {166711} (\bibinfo
		{year} {2020})}\BibitemShut {NoStop}%
	\bibitem [{\citenamefont {Shao}\ \emph {et~al.}(2021)\citenamefont {Shao},
		\citenamefont {Li}, \citenamefont {Liu}, \citenamefont {Yang}, \citenamefont
		{Fukami}, \citenamefont {Razavi}, \citenamefont {Wu}, \citenamefont {Wang},
		\citenamefont {Freimuth}, \citenamefont {Mokrousov}, \citenamefont {Stiles},
		\citenamefont {Emori}, \citenamefont {Hoffmann}, \citenamefont {Åkerman},
		\citenamefont {Roy}, \citenamefont {Wang}, \citenamefont {Yang},
		\citenamefont {Garello},\ and\ \citenamefont {Zhang}}]{Roadmap-SOT-Review}%
	\BibitemOpen
	\bibfield  {author} {\bibinfo {author} {\bibfnamefont {Q.}~\bibnamefont
			{Shao}}, \bibinfo {author} {\bibfnamefont {P.}~\bibnamefont {Li}}, \bibinfo
		{author} {\bibfnamefont {L.}~\bibnamefont {Liu}}, \bibinfo {author}
		{\bibfnamefont {H.}~\bibnamefont {Yang}}, \bibinfo {author} {\bibfnamefont
			{S.}~\bibnamefont {Fukami}}, \bibinfo {author} {\bibfnamefont
			{A.}~\bibnamefont {Razavi}}, \bibinfo {author} {\bibfnamefont
			{H.}~\bibnamefont {Wu}}, \bibinfo {author} {\bibfnamefont {K.}~\bibnamefont
			{Wang}}, \bibinfo {author} {\bibfnamefont {F.}~\bibnamefont {Freimuth}},
		\bibinfo {author} {\bibfnamefont {Y.}~\bibnamefont {Mokrousov}}, \bibinfo
		{author} {\bibfnamefont {M.~D.}\ \bibnamefont {Stiles}}, \bibinfo {author}
		{\bibfnamefont {S.}~\bibnamefont {Emori}}, \bibinfo {author} {\bibfnamefont
			{A.}~\bibnamefont {Hoffmann}}, \bibinfo {author} {\bibfnamefont
			{J.}~\bibnamefont {Åkerman}}, \bibinfo {author} {\bibfnamefont
			{K.}~\bibnamefont {Roy}}, \bibinfo {author} {\bibfnamefont {J.-P.}\
			\bibnamefont {Wang}}, \bibinfo {author} {\bibfnamefont {S.-H.}\ \bibnamefont
			{Yang}}, \bibinfo {author} {\bibfnamefont {K.}~\bibnamefont {Garello}},\ and\
		\bibinfo {author} {\bibfnamefont {W.}~\bibnamefont {Zhang}},\ }\href
	{https://doi.org/10.1109/TMAG.2021.3078583} {\bibfield  {journal} {\bibinfo
			{journal} {IEEE Transactions on Magnetics}\ }\textbf {\bibinfo {volume}
			{57}},\ \bibinfo {pages} {1} (\bibinfo {year} {2021})}\BibitemShut {NoStop}%
	\bibitem [{\citenamefont {Choi}\ \emph {et~al.}(2023)\citenamefont {Choi},
		\citenamefont {Jo}, \citenamefont {Ko}, \citenamefont {Go}, \citenamefont
		{Kim}, \citenamefont {Park}, \citenamefont {Kim}, \citenamefont {Min},
		\citenamefont {Choi},\ and\ \citenamefont
		{Lee}}]{Exp-OHE-Ti-Nat-2023-Hyun-Woo}%
	\BibitemOpen
	\bibfield  {author} {\bibinfo {author} {\bibfnamefont {Y.-G.}\ \bibnamefont
			{Choi}}, \bibinfo {author} {\bibfnamefont {D.}~\bibnamefont {Jo}}, \bibinfo
		{author} {\bibfnamefont {K.-H.}\ \bibnamefont {Ko}}, \bibinfo {author}
		{\bibfnamefont {D.}~\bibnamefont {Go}}, \bibinfo {author} {\bibfnamefont
			{K.-H.}\ \bibnamefont {Kim}}, \bibinfo {author} {\bibfnamefont {H.~G.}\
			\bibnamefont {Park}}, \bibinfo {author} {\bibfnamefont {C.}~\bibnamefont
			{Kim}}, \bibinfo {author} {\bibfnamefont {B.-C.}\ \bibnamefont {Min}},
		\bibinfo {author} {\bibfnamefont {G.-M.}\ \bibnamefont {Choi}},\ and\
		\bibinfo {author} {\bibfnamefont {H.-W.}\ \bibnamefont {Lee}},\ }\href
	{https://doi.org/10.1038/s41586-023-06101-9} {\bibfield  {journal} {\bibinfo
			{journal} {Nature}\ }\textbf {\bibinfo {volume} {619}},\ \bibinfo {pages}
		{52} (\bibinfo {year} {2023})}\BibitemShut {NoStop}%
	\bibitem [{\citenamefont {Liu}\ and\ \citenamefont
		{Culcer}(2024)}]{Hong-OHE-PRL}%
	\BibitemOpen
	\bibfield  {author} {\bibinfo {author} {\bibfnamefont {H.}~\bibnamefont
			{Liu}}\ and\ \bibinfo {author} {\bibfnamefont {D.}~\bibnamefont {Culcer}},\
	}\href {https://doi.org/10.1103/PhysRevLett.132.186302} {\bibfield  {journal}
		{\bibinfo  {journal} {Phys. Rev. Lett.}\ }\textbf {\bibinfo {volume} {132}},\
		\bibinfo {pages} {186302} (\bibinfo {year} {2024})}\BibitemShut {NoStop}%
	\bibitem [{\citenamefont {Xiao}\ \emph {et~al.}()\citenamefont {Xiao},
		\citenamefont {Liu},\ and\ \citenamefont {Yan}}]{OHE-Binghai}%
	\BibitemOpen
	\bibfield  {author} {\bibinfo {author} {\bibfnamefont {J.}~\bibnamefont
			{Xiao}}, \bibinfo {author} {\bibfnamefont {Y.}~\bibnamefont {Liu}},\ and\
		\bibinfo {author} {\bibfnamefont {B.}~\bibnamefont {Yan}},\ }\bibinfo {title}
	{Detection of the orbital hall effect by the orbital--spin conversion},\ in\
	\href {https://doi.org/10.1142/9789811231711_0015} {\emph {\bibinfo
			{booktitle} {Memorial Volume for Shoucheng Zhang}}},\ Chap.\ \bibinfo
	{chapter} {Chapter 13}, pp.\ \bibinfo {pages} {353--364}\BibitemShut
	{NoStop}%
	\bibitem [{\citenamefont {Pezo}\ \emph {et~al.}(2022)\citenamefont {Pezo},
		\citenamefont {Garc\'{\i}a~Ovalle},\ and\ \citenamefont
		{Manchon}}]{OHE-PRB-2022-Manchon}%
	\BibitemOpen
	\bibfield  {author} {\bibinfo {author} {\bibfnamefont {A.}~\bibnamefont
			{Pezo}}, \bibinfo {author} {\bibfnamefont {D.}~\bibnamefont
			{Garc\'{\i}a~Ovalle}},\ and\ \bibinfo {author} {\bibfnamefont
			{A.}~\bibnamefont {Manchon}},\ }\href
	{https://doi.org/10.1103/PhysRevB.106.104414} {\bibfield  {journal} {\bibinfo
			{journal} {Phys. Rev. B}\ }\textbf {\bibinfo {volume} {106}},\ \bibinfo
		{pages} {104414} (\bibinfo {year} {2022})}\BibitemShut {NoStop}%
	\bibitem [{\citenamefont {Salemi}\ and\ \citenamefont
		{Oppeneer}(2022)}]{OHE-metal-PRM-2022-Oppeneer}%
	\BibitemOpen
	\bibfield  {author} {\bibinfo {author} {\bibfnamefont {L.}~\bibnamefont
			{Salemi}}\ and\ \bibinfo {author} {\bibfnamefont {P.~M.}\ \bibnamefont
			{Oppeneer}},\ }\href {https://doi.org/10.1103/PhysRevMaterials.6.095001}
	{\bibfield  {journal} {\bibinfo  {journal} {Phys. Rev. Mater.}\ }\textbf
		{\bibinfo {volume} {6}},\ \bibinfo {pages} {095001} (\bibinfo {year}
		{2022})}\BibitemShut {NoStop}%
	\bibitem [{\citenamefont {Cysne}\ \emph {et~al.}(2021)\citenamefont {Cysne},
		\citenamefont {Costa}, \citenamefont {Canonico}, \citenamefont {Nardelli},
		\citenamefont {Muniz},\ and\ \citenamefont
		{Rappoport}}]{OHE-BiTMD-PRL-2021-Tatiana}%
	\BibitemOpen
	\bibfield  {author} {\bibinfo {author} {\bibfnamefont {T.~P.}\ \bibnamefont
			{Cysne}}, \bibinfo {author} {\bibfnamefont {M.}~\bibnamefont {Costa}},
		\bibinfo {author} {\bibfnamefont {L.~M.}\ \bibnamefont {Canonico}}, \bibinfo
		{author} {\bibfnamefont {M.~B.}\ \bibnamefont {Nardelli}}, \bibinfo {author}
		{\bibfnamefont {R.~B.}\ \bibnamefont {Muniz}},\ and\ \bibinfo {author}
		{\bibfnamefont {T.~G.}\ \bibnamefont {Rappoport}},\ }\href
	{https://doi.org/10.1103/PhysRevLett.126.056601} {\bibfield  {journal}
		{\bibinfo  {journal} {Phys. Rev. Lett.}\ }\textbf {\bibinfo {volume} {126}},\
		\bibinfo {pages} {056601} (\bibinfo {year} {2021})}\BibitemShut {NoStop}%
	\bibitem [{\citenamefont {Luis M.~Canonico}(2024)}]{RS-OHE-disorder}%
	\BibitemOpen
	\bibfield  {author} {\bibinfo {author} {\bibfnamefont {S.~R.}\ \bibnamefont
			{Luis M.~Canonico}, \bibfnamefont {Jose H.~García}},\ }\href@noop {}
	{\bibfield  {journal} {\bibinfo  {journal} {arXiv:2404.01739}\ } (\bibinfo
		{year} {2024})}\BibitemShut {NoStop}%
	\bibitem [{\citenamefont {Go}\ \emph {et~al.}(2018{\natexlab{a}})\citenamefont
		{Go}, \citenamefont {Jo}, \citenamefont {Kim},\ and\ \citenamefont
		{Lee}}]{ISOHE-PRL-2018-Hyun-Woo}%
	\BibitemOpen
	\bibfield  {author} {\bibinfo {author} {\bibfnamefont {D.}~\bibnamefont
			{Go}}, \bibinfo {author} {\bibfnamefont {D.}~\bibnamefont {Jo}}, \bibinfo
		{author} {\bibfnamefont {C.}~\bibnamefont {Kim}},\ and\ \bibinfo {author}
		{\bibfnamefont {H.-W.}\ \bibnamefont {Lee}},\ }\href
	{https://doi.org/10.1103/PhysRevLett.121.086602} {\bibfield  {journal}
		{\bibinfo  {journal} {Phys. Rev. Lett.}\ }\textbf {\bibinfo {volume} {121}},\
		\bibinfo {pages} {086602} (\bibinfo {year} {2018}{\natexlab{a}})}\BibitemShut
	{NoStop}%
	\bibitem [{\citenamefont {Jo}\ \emph {et~al.}(2018)\citenamefont {Jo},
		\citenamefont {Go},\ and\ \citenamefont
		{Lee}}]{IOHE-Metal-PRB-2018-Hyun-Woo}%
	\BibitemOpen
	\bibfield  {author} {\bibinfo {author} {\bibfnamefont {D.}~\bibnamefont
			{Jo}}, \bibinfo {author} {\bibfnamefont {D.}~\bibnamefont {Go}},\ and\
		\bibinfo {author} {\bibfnamefont {H.-W.}\ \bibnamefont {Lee}},\ }\href
	{https://doi.org/10.1103/PhysRevB.98.214405} {\bibfield  {journal} {\bibinfo
			{journal} {Phys. Rev. B}\ }\textbf {\bibinfo {volume} {98}},\ \bibinfo
		{pages} {214405} (\bibinfo {year} {2018})}\BibitemShut {NoStop}%
	\bibitem [{\citenamefont {Sala}\ and\ \citenamefont
		{Gambardella}(2022{\natexlab{a}})}]{OHE-Hetero-PRR-2022-Pietro}%
	\BibitemOpen
	\bibfield  {author} {\bibinfo {author} {\bibfnamefont {G.}~\bibnamefont
			{Sala}}\ and\ \bibinfo {author} {\bibfnamefont {P.}~\bibnamefont
			{Gambardella}},\ }\href {https://doi.org/10.1103/PhysRevResearch.4.033037}
	{\bibfield  {journal} {\bibinfo  {journal} {Phys. Rev. Res.}\ }\textbf
		{\bibinfo {volume} {4}},\ \bibinfo {pages} {033037} (\bibinfo {year}
		{2022}{\natexlab{a}})}\BibitemShut {NoStop}%
	\bibitem [{\citenamefont {Wang}\ \emph
		{et~al.}(2023{\natexlab{a}})\citenamefont {Wang}, \citenamefont {Feng},
		\citenamefont {Yang}, \citenamefont {Zhang}, \citenamefont {Liu},
		\citenamefont {Xu}, \citenamefont {Jia}, \citenamefont {Wu}, \citenamefont
		{Yu}, \citenamefont {Xu},\ and\ \citenamefont {Jiang}}]{OHE-Weak-SOC-npj}%
	\BibitemOpen
	\bibfield  {author} {\bibinfo {author} {\bibfnamefont {P.}~\bibnamefont
			{Wang}}, \bibinfo {author} {\bibfnamefont {Z.}~\bibnamefont {Feng}}, \bibinfo
		{author} {\bibfnamefont {Y.}~\bibnamefont {Yang}}, \bibinfo {author}
		{\bibfnamefont {D.}~\bibnamefont {Zhang}}, \bibinfo {author} {\bibfnamefont
			{Q.}~\bibnamefont {Liu}}, \bibinfo {author} {\bibfnamefont {Z.}~\bibnamefont
			{Xu}}, \bibinfo {author} {\bibfnamefont {Z.}~\bibnamefont {Jia}}, \bibinfo
		{author} {\bibfnamefont {Y.}~\bibnamefont {Wu}}, \bibinfo {author}
		{\bibfnamefont {G.}~\bibnamefont {Yu}}, \bibinfo {author} {\bibfnamefont
			{X.}~\bibnamefont {Xu}},\ and\ \bibinfo {author} {\bibfnamefont
			{Y.}~\bibnamefont {Jiang}},\ }\href
	{https://doi.org/10.1038/s41535-023-00559-6} {\bibfield  {journal} {\bibinfo
			{journal} {npj Quantum Materials}\ }\textbf {\bibinfo {volume} {8}},\
		\bibinfo {pages} {28} (\bibinfo {year} {2023}{\natexlab{a}})}\BibitemShut
	{NoStop}%
	\bibitem [{\citenamefont {Lyalin}\ \emph {et~al.}(2023)\citenamefont {Lyalin},
		\citenamefont {Alikhah}, \citenamefont {Berritta}, \citenamefont {Oppeneer},\
		and\ \citenamefont {Kawakami}}]{PhysRevLett.131.156702}%
	\BibitemOpen
	\bibfield  {author} {\bibinfo {author} {\bibfnamefont {I.}~\bibnamefont
			{Lyalin}}, \bibinfo {author} {\bibfnamefont {S.}~\bibnamefont {Alikhah}},
		\bibinfo {author} {\bibfnamefont {M.}~\bibnamefont {Berritta}}, \bibinfo
		{author} {\bibfnamefont {P.~M.}\ \bibnamefont {Oppeneer}},\ and\ \bibinfo
		{author} {\bibfnamefont {R.~K.}\ \bibnamefont {Kawakami}},\ }\href
	{https://doi.org/10.1103/PhysRevLett.131.156702} {\bibfield  {journal}
		{\bibinfo  {journal} {Phys. Rev. Lett.}\ }\textbf {\bibinfo {volume} {131}},\
		\bibinfo {pages} {156702} (\bibinfo {year} {2023})}\BibitemShut {NoStop}%
	\bibitem [{\citenamefont {Dutta}\ and\ \citenamefont
		{Tulapurkar}(2022)}]{PhysRevB.106.184406}%
	\BibitemOpen
	\bibfield  {author} {\bibinfo {author} {\bibfnamefont {S.}~\bibnamefont
			{Dutta}}\ and\ \bibinfo {author} {\bibfnamefont {A.~A.}\ \bibnamefont
			{Tulapurkar}},\ }\href {https://doi.org/10.1103/PhysRevB.106.184406}
	{\bibfield  {journal} {\bibinfo  {journal} {Phys. Rev. B}\ }\textbf {\bibinfo
			{volume} {106}},\ \bibinfo {pages} {184406} (\bibinfo {year}
		{2022})}\BibitemShut {NoStop}%
	\bibitem [{\citenamefont {Sala}\ \emph {et~al.}(2023)\citenamefont {Sala},
		\citenamefont {Wang}, \citenamefont {Legrand},\ and\ \citenamefont
		{Gambardella}}]{PhysRevLett.131.156703}%
	\BibitemOpen
	\bibfield  {author} {\bibinfo {author} {\bibfnamefont {G.}~\bibnamefont
			{Sala}}, \bibinfo {author} {\bibfnamefont {H.}~\bibnamefont {Wang}}, \bibinfo
		{author} {\bibfnamefont {W.}~\bibnamefont {Legrand}},\ and\ \bibinfo {author}
		{\bibfnamefont {P.}~\bibnamefont {Gambardella}},\ }\href
	{https://doi.org/10.1103/PhysRevLett.131.156703} {\bibfield  {journal}
		{\bibinfo  {journal} {Phys. Rev. Lett.}\ }\textbf {\bibinfo {volume} {131}},\
		\bibinfo {pages} {156703} (\bibinfo {year} {2023})}\BibitemShut {NoStop}%
	\bibitem [{\citenamefont {Zhang}\ \emph {et~al.}(2022)\citenamefont {Zhang},
		\citenamefont {Xie}, \citenamefont {Zhang}, \citenamefont {Yan},
		\citenamefont {Zhai}, \citenamefont {Chi}, \citenamefont {Xu}, \citenamefont
		{Zuo},\ and\ \citenamefont {Xi}}]{10.1063/5.0106988}%
	\BibitemOpen
	\bibfield  {author} {\bibinfo {author} {\bibfnamefont {J.}~\bibnamefont
			{Zhang}}, \bibinfo {author} {\bibfnamefont {H.}~\bibnamefont {Xie}}, \bibinfo
		{author} {\bibfnamefont {X.}~\bibnamefont {Zhang}}, \bibinfo {author}
		{\bibfnamefont {Z.}~\bibnamefont {Yan}}, \bibinfo {author} {\bibfnamefont
			{Y.}~\bibnamefont {Zhai}}, \bibinfo {author} {\bibfnamefont {J.}~\bibnamefont
			{Chi}}, \bibinfo {author} {\bibfnamefont {H.}~\bibnamefont {Xu}}, \bibinfo
		{author} {\bibfnamefont {Y.}~\bibnamefont {Zuo}},\ and\ \bibinfo {author}
		{\bibfnamefont {L.}~\bibnamefont {Xi}},\ }\href
	{https://doi.org/10.1063/5.0106988} {\bibfield  {journal} {\bibinfo
			{journal} {Applied Physics Letters}\ }\textbf {\bibinfo {volume} {121}},\
		\bibinfo {pages} {172405} (\bibinfo {year} {2022})},\ \Eprint
	{https://arxiv.org/abs/https://pubs.aip.org/aip/apl/article-pdf/doi/10.1063/5.0106988/16485598/172405\_1\_online.pdf}
	{https://pubs.aip.org/aip/apl/article-pdf/doi/10.1063/5.0106988/16485598/172405\_1\_online.pdf}
	\BibitemShut {NoStop}%
	\bibitem [{\citenamefont {Ding}\ \emph {et~al.}(2022)\citenamefont {Ding},
		\citenamefont {Liang}, \citenamefont {Go}, \citenamefont {Yun}, \citenamefont
		{Xue}, \citenamefont {Liu}, \citenamefont {Becker}, \citenamefont {Yang},
		\citenamefont {Du}, \citenamefont {Wang}, \citenamefont {Yang}, \citenamefont
		{Jakob}, \citenamefont {Kl\"aui}, \citenamefont {Mokrousov},\ and\
		\citenamefont {Yang}}]{Exp-OEE-PRL-2022-Jinbo}%
	\BibitemOpen
	\bibfield  {author} {\bibinfo {author} {\bibfnamefont {S.}~\bibnamefont
			{Ding}}, \bibinfo {author} {\bibfnamefont {Z.}~\bibnamefont {Liang}},
		\bibinfo {author} {\bibfnamefont {D.}~\bibnamefont {Go}}, \bibinfo {author}
		{\bibfnamefont {C.}~\bibnamefont {Yun}}, \bibinfo {author} {\bibfnamefont
			{M.}~\bibnamefont {Xue}}, \bibinfo {author} {\bibfnamefont {Z.}~\bibnamefont
			{Liu}}, \bibinfo {author} {\bibfnamefont {S.}~\bibnamefont {Becker}},
		\bibinfo {author} {\bibfnamefont {W.}~\bibnamefont {Yang}}, \bibinfo {author}
		{\bibfnamefont {H.}~\bibnamefont {Du}}, \bibinfo {author} {\bibfnamefont
			{C.}~\bibnamefont {Wang}}, \bibinfo {author} {\bibfnamefont {Y.}~\bibnamefont
			{Yang}}, \bibinfo {author} {\bibfnamefont {G.}~\bibnamefont {Jakob}},
		\bibinfo {author} {\bibfnamefont {M.}~\bibnamefont {Kl\"aui}}, \bibinfo
		{author} {\bibfnamefont {Y.}~\bibnamefont {Mokrousov}},\ and\ \bibinfo
		{author} {\bibfnamefont {J.}~\bibnamefont {Yang}},\ }\href
	{https://doi.org/10.1103/PhysRevLett.128.067201} {\bibfield  {journal}
		{\bibinfo  {journal} {Phys. Rev. Lett.}\ }\textbf {\bibinfo {volume} {128}},\
		\bibinfo {pages} {067201} (\bibinfo {year} {2022})}\BibitemShut {NoStop}%
	\bibitem [{\citenamefont {Voss}\ \emph {et~al.}(2024)\citenamefont {Voss},
		\citenamefont {Ado},\ and\ \citenamefont {Titov}}]{Titov_EdgeOM}%
	\BibitemOpen
	\bibfield  {author} {\bibinfo {author} {\bibfnamefont {J.}~\bibnamefont
			{Voss}}, \bibinfo {author} {\bibfnamefont {I.}~\bibnamefont {Ado}},\ and\
		\bibinfo {author} {\bibfnamefont {M.}~\bibnamefont {Titov}},\ }\href@noop {}
	{\bibfield  {journal} {\bibinfo  {journal} {arXiv preprint arXiv:2405.11979}\
		} (\bibinfo {year} {2024})}\BibitemShut {NoStop}%
	\bibitem [{\citenamefont {Kim}\ \emph {et~al.}(2021)\citenamefont {Kim},
		\citenamefont {Go}, \citenamefont {Tsai}, \citenamefont {Jo}, \citenamefont
		{Kondou}, \citenamefont {Lee},\ and\ \citenamefont
		{Otani}}]{OT-FM-PRB-2021-YoshiChika}%
	\BibitemOpen
	\bibfield  {author} {\bibinfo {author} {\bibfnamefont {J.}~\bibnamefont
			{Kim}}, \bibinfo {author} {\bibfnamefont {D.}~\bibnamefont {Go}}, \bibinfo
		{author} {\bibfnamefont {H.}~\bibnamefont {Tsai}}, \bibinfo {author}
		{\bibfnamefont {D.}~\bibnamefont {Jo}}, \bibinfo {author} {\bibfnamefont
			{K.}~\bibnamefont {Kondou}}, \bibinfo {author} {\bibfnamefont {H.-W.}\
			\bibnamefont {Lee}},\ and\ \bibinfo {author} {\bibfnamefont {Y.}~\bibnamefont
			{Otani}},\ }\href {https://doi.org/10.1103/PhysRevB.103.L020407} {\bibfield
		{journal} {\bibinfo  {journal} {Phys. Rev. B}\ }\textbf {\bibinfo {volume}
			{103}},\ \bibinfo {pages} {L020407} (\bibinfo {year} {2021})}\BibitemShut
	{NoStop}%
	\bibitem [{\citenamefont {Chen}\ \emph {et~al.}(2018)\citenamefont {Chen},
		\citenamefont {Liu}, \citenamefont {Yang}, \citenamefont {Shi}, \citenamefont
		{Hu}, \citenamefont {Li},\ and\ \citenamefont
		{Zeng}}]{OT-OEE-NatComm-2018-Haibo}%
	\BibitemOpen
	\bibfield  {author} {\bibinfo {author} {\bibfnamefont {X.}~\bibnamefont
			{Chen}}, \bibinfo {author} {\bibfnamefont {Y.}~\bibnamefont {Liu}}, \bibinfo
		{author} {\bibfnamefont {G.}~\bibnamefont {Yang}}, \bibinfo {author}
		{\bibfnamefont {H.}~\bibnamefont {Shi}}, \bibinfo {author} {\bibfnamefont
			{C.}~\bibnamefont {Hu}}, \bibinfo {author} {\bibfnamefont {M.}~\bibnamefont
			{Li}},\ and\ \bibinfo {author} {\bibfnamefont {H.}~\bibnamefont {Zeng}},\
	}\href {https://doi.org/10.1038/s41467-018-05057-z} {\bibfield  {journal}
		{\bibinfo  {journal} {Nature Communications}\ }\textbf {\bibinfo {volume}
			{9}},\ \bibinfo {pages} {2569} (\bibinfo {year} {2018})}\BibitemShut
	{NoStop}%
	\bibitem [{\citenamefont {Go}\ and\ \citenamefont
		{Lee}(2020)}]{OT-PRR-2020-Hyun-Woo}%
	\BibitemOpen
	\bibfield  {author} {\bibinfo {author} {\bibfnamefont {D.}~\bibnamefont
			{Go}}\ and\ \bibinfo {author} {\bibfnamefont {H.-W.}\ \bibnamefont {Lee}},\
	}\href {https://doi.org/10.1103/PhysRevResearch.2.013177} {\bibfield
		{journal} {\bibinfo  {journal} {Phys. Rev. Res.}\ }\textbf {\bibinfo {volume}
			{2}},\ \bibinfo {pages} {013177} (\bibinfo {year} {2020})}\BibitemShut
	{NoStop}%
	\bibitem [{\citenamefont {Lee}\ \emph {et~al.}(2021{\natexlab{a}})\citenamefont
		{Lee}, \citenamefont {Go}, \citenamefont {Park}, \citenamefont {Jeong},
		\citenamefont {Ko}, \citenamefont {Yun}, \citenamefont {Jo}, \citenamefont
		{Lee}, \citenamefont {Go}, \citenamefont {Oh}, \citenamefont {Kim},
		\citenamefont {Park}, \citenamefont {Min}, \citenamefont {Koo}, \citenamefont
		{Lee}, \citenamefont {Lee},\ and\ \citenamefont
		{Lee}}]{OT-NatComm-2021-Kyung-Jin}%
	\BibitemOpen
	\bibfield  {author} {\bibinfo {author} {\bibfnamefont {D.}~\bibnamefont
			{Lee}}, \bibinfo {author} {\bibfnamefont {D.}~\bibnamefont {Go}}, \bibinfo
		{author} {\bibfnamefont {H.-J.}\ \bibnamefont {Park}}, \bibinfo {author}
		{\bibfnamefont {W.}~\bibnamefont {Jeong}}, \bibinfo {author} {\bibfnamefont
			{H.-W.}\ \bibnamefont {Ko}}, \bibinfo {author} {\bibfnamefont
			{D.}~\bibnamefont {Yun}}, \bibinfo {author} {\bibfnamefont {D.}~\bibnamefont
			{Jo}}, \bibinfo {author} {\bibfnamefont {S.}~\bibnamefont {Lee}}, \bibinfo
		{author} {\bibfnamefont {G.}~\bibnamefont {Go}}, \bibinfo {author}
		{\bibfnamefont {J.~H.}\ \bibnamefont {Oh}}, \bibinfo {author} {\bibfnamefont
			{K.-J.}\ \bibnamefont {Kim}}, \bibinfo {author} {\bibfnamefont {B.-G.}\
			\bibnamefont {Park}}, \bibinfo {author} {\bibfnamefont {B.-C.}\ \bibnamefont
			{Min}}, \bibinfo {author} {\bibfnamefont {H.~C.}\ \bibnamefont {Koo}},
		\bibinfo {author} {\bibfnamefont {H.-W.}\ \bibnamefont {Lee}}, \bibinfo
		{author} {\bibfnamefont {O.}~\bibnamefont {Lee}},\ and\ \bibinfo {author}
		{\bibfnamefont {K.-J.}\ \bibnamefont {Lee}},\ }\href
	{https://doi.org/10.1038/s41467-021-26650-9} {\bibfield  {journal} {\bibinfo
			{journal} {Nature Communications}\ }\textbf {\bibinfo {volume} {12}},\
		\bibinfo {pages} {6710} (\bibinfo {year} {2021}{\natexlab{a}})}\BibitemShut
	{NoStop}%
	\bibitem [{\citenamefont {Zheng}\ \emph {et~al.}(2020)\citenamefont {Zheng},
		\citenamefont {Guo}, \citenamefont {Jo}, \citenamefont {Go}, \citenamefont
		{Wang}, \citenamefont {Chen}, \citenamefont {Yin}, \citenamefont {Wang},
		\citenamefont {Yu}, \citenamefont {He}, \citenamefont {Lee}, \citenamefont
		{Teng},\ and\ \citenamefont {Zhu}}]{Exp-OT-PRR-2020}%
	\BibitemOpen
	\bibfield  {author} {\bibinfo {author} {\bibfnamefont {Z.~C.}\ \bibnamefont
			{Zheng}}, \bibinfo {author} {\bibfnamefont {Q.~X.}\ \bibnamefont {Guo}},
		\bibinfo {author} {\bibfnamefont {D.}~\bibnamefont {Jo}}, \bibinfo {author}
		{\bibfnamefont {D.}~\bibnamefont {Go}}, \bibinfo {author} {\bibfnamefont
			{L.~H.}\ \bibnamefont {Wang}}, \bibinfo {author} {\bibfnamefont {H.~C.}\
			\bibnamefont {Chen}}, \bibinfo {author} {\bibfnamefont {W.}~\bibnamefont
			{Yin}}, \bibinfo {author} {\bibfnamefont {X.~M.}\ \bibnamefont {Wang}},
		\bibinfo {author} {\bibfnamefont {G.~H.}\ \bibnamefont {Yu}}, \bibinfo
		{author} {\bibfnamefont {W.}~\bibnamefont {He}}, \bibinfo {author}
		{\bibfnamefont {H.-W.}\ \bibnamefont {Lee}}, \bibinfo {author} {\bibfnamefont
			{J.}~\bibnamefont {Teng}},\ and\ \bibinfo {author} {\bibfnamefont
			{T.}~\bibnamefont {Zhu}},\ }\href
	{https://doi.org/10.1103/PhysRevResearch.2.013127} {\bibfield  {journal}
		{\bibinfo  {journal} {Phys. Rev. Res.}\ }\textbf {\bibinfo {volume} {2}},\
		\bibinfo {pages} {013127} (\bibinfo {year} {2020})}\BibitemShut {NoStop}%
	\bibitem [{\citenamefont {Lee}\ \emph {et~al.}(2021{\natexlab{b}})\citenamefont
		{Lee}, \citenamefont {Kang}, \citenamefont {Go}, \citenamefont {Kim},
		\citenamefont {Kang}, \citenamefont {Lee}, \citenamefont {Lee}, \citenamefont
		{Kang}, \citenamefont {Lee}, \citenamefont {Mokrousov}, \citenamefont {Kim},
		\citenamefont {Kim}, \citenamefont {Lee},\ and\ \citenamefont
		{Park}}]{Exp-OT-CommP-2021-Byong-Guk}%
	\BibitemOpen
	\bibfield  {author} {\bibinfo {author} {\bibfnamefont {S.}~\bibnamefont
			{Lee}}, \bibinfo {author} {\bibfnamefont {M.-G.}\ \bibnamefont {Kang}},
		\bibinfo {author} {\bibfnamefont {D.}~\bibnamefont {Go}}, \bibinfo {author}
		{\bibfnamefont {D.}~\bibnamefont {Kim}}, \bibinfo {author} {\bibfnamefont
			{J.-H.}\ \bibnamefont {Kang}}, \bibinfo {author} {\bibfnamefont
			{T.}~\bibnamefont {Lee}}, \bibinfo {author} {\bibfnamefont {G.-H.}\
			\bibnamefont {Lee}}, \bibinfo {author} {\bibfnamefont {J.}~\bibnamefont
			{Kang}}, \bibinfo {author} {\bibfnamefont {N.~J.}\ \bibnamefont {Lee}},
		\bibinfo {author} {\bibfnamefont {Y.}~\bibnamefont {Mokrousov}}, \bibinfo
		{author} {\bibfnamefont {S.}~\bibnamefont {Kim}}, \bibinfo {author}
		{\bibfnamefont {K.-J.}\ \bibnamefont {Kim}}, \bibinfo {author} {\bibfnamefont
			{K.-J.}\ \bibnamefont {Lee}},\ and\ \bibinfo {author} {\bibfnamefont {B.-G.}\
			\bibnamefont {Park}},\ }\href {https://doi.org/10.1038/s42005-021-00737-7}
	{\bibfield  {journal} {\bibinfo  {journal} {Communications Physics}\ }\textbf
		{\bibinfo {volume} {4}},\ \bibinfo {pages} {234} (\bibinfo {year}
		{2021}{\natexlab{b}})}\BibitemShut {NoStop}%
	\bibitem [{\citenamefont {Lee}\ \emph {et~al.}(2021{\natexlab{c}})\citenamefont
		{Lee}, \citenamefont {Kang}, \citenamefont {Go}, \citenamefont {Kim},
		\citenamefont {Kang}, \citenamefont {Lee}, \citenamefont {Lee}, \citenamefont
		{Kang}, \citenamefont {Lee}, \citenamefont {Mokrousov}, \citenamefont {Kim},
		\citenamefont {Kim}, \citenamefont {Lee},\ and\ \citenamefont
		{Park}}]{LS-conversion-CP-2021-Byong-Guk}%
	\BibitemOpen
	\bibfield  {author} {\bibinfo {author} {\bibfnamefont {S.}~\bibnamefont
			{Lee}}, \bibinfo {author} {\bibfnamefont {M.-G.}\ \bibnamefont {Kang}},
		\bibinfo {author} {\bibfnamefont {D.}~\bibnamefont {Go}}, \bibinfo {author}
		{\bibfnamefont {D.}~\bibnamefont {Kim}}, \bibinfo {author} {\bibfnamefont
			{J.-H.}\ \bibnamefont {Kang}}, \bibinfo {author} {\bibfnamefont
			{T.}~\bibnamefont {Lee}}, \bibinfo {author} {\bibfnamefont {G.-H.}\
			\bibnamefont {Lee}}, \bibinfo {author} {\bibfnamefont {J.}~\bibnamefont
			{Kang}}, \bibinfo {author} {\bibfnamefont {N.~J.}\ \bibnamefont {Lee}},
		\bibinfo {author} {\bibfnamefont {Y.}~\bibnamefont {Mokrousov}}, \bibinfo
		{author} {\bibfnamefont {S.}~\bibnamefont {Kim}}, \bibinfo {author}
		{\bibfnamefont {K.-J.}\ \bibnamefont {Kim}}, \bibinfo {author} {\bibfnamefont
			{K.-J.}\ \bibnamefont {Lee}},\ and\ \bibinfo {author} {\bibfnamefont {B.-G.}\
			\bibnamefont {Park}},\ }\href@noop {} {\bibfield  {journal} {\bibinfo
			{journal} {Communications Physics}\ }\textbf {\bibinfo {volume} {4}},\
		\bibinfo {pages} {234} (\bibinfo {year} {2021}{\natexlab{c}})}\BibitemShut
	{NoStop}%
	\bibitem [{\citenamefont {Ding}\ \emph {et~al.}(2020)\citenamefont {Ding},
		\citenamefont {Ross}, \citenamefont {Go}, \citenamefont {Baldrati},
		\citenamefont {Ren}, \citenamefont {Freimuth}, \citenamefont {Becker},
		\citenamefont {Kammerbauer}, \citenamefont {Yang}, \citenamefont {Jakob},
		\citenamefont {Mokrousov},\ and\ \citenamefont
		{Kl\"aui}}]{OOS-Cvert-2020-PRL-Mathias}%
	\BibitemOpen
	\bibfield  {author} {\bibinfo {author} {\bibfnamefont {S.}~\bibnamefont
			{Ding}}, \bibinfo {author} {\bibfnamefont {A.}~\bibnamefont {Ross}}, \bibinfo
		{author} {\bibfnamefont {D.}~\bibnamefont {Go}}, \bibinfo {author}
		{\bibfnamefont {L.}~\bibnamefont {Baldrati}}, \bibinfo {author}
		{\bibfnamefont {Z.}~\bibnamefont {Ren}}, \bibinfo {author} {\bibfnamefont
			{F.}~\bibnamefont {Freimuth}}, \bibinfo {author} {\bibfnamefont
			{S.}~\bibnamefont {Becker}}, \bibinfo {author} {\bibfnamefont
			{F.}~\bibnamefont {Kammerbauer}}, \bibinfo {author} {\bibfnamefont
			{J.}~\bibnamefont {Yang}}, \bibinfo {author} {\bibfnamefont {G.}~\bibnamefont
			{Jakob}}, \bibinfo {author} {\bibfnamefont {Y.}~\bibnamefont {Mokrousov}},\
		and\ \bibinfo {author} {\bibfnamefont {M.}~\bibnamefont {Kl\"aui}},\ }\href
	{https://doi.org/10.1103/PhysRevLett.125.177201} {\bibfield  {journal}
		{\bibinfo  {journal} {Phys. Rev. Lett.}\ }\textbf {\bibinfo {volume} {125}},\
		\bibinfo {pages} {177201} (\bibinfo {year} {2020})}\BibitemShut {NoStop}%
	\bibitem [{\citenamefont {Sala}\ and\ \citenamefont
		{Gambardella}(2022{\natexlab{b}})}]{PhysRevResearch.4.033037}%
	\BibitemOpen
	\bibfield  {author} {\bibinfo {author} {\bibfnamefont {G.}~\bibnamefont
			{Sala}}\ and\ \bibinfo {author} {\bibfnamefont {P.}~\bibnamefont
			{Gambardella}},\ }\href {https://doi.org/10.1103/PhysRevResearch.4.033037}
	{\bibfield  {journal} {\bibinfo  {journal} {Phys. Rev. Res.}\ }\textbf
		{\bibinfo {volume} {4}},\ \bibinfo {pages} {033037} (\bibinfo {year}
		{2022}{\natexlab{b}})}\BibitemShut {NoStop}%
	\bibitem [{\citenamefont {Hayashi}\ \emph {et~al.}(2023)\citenamefont
		{Hayashi}, \citenamefont {Jo}, \citenamefont {Go}, \citenamefont {Gao},
		\citenamefont {Haku}, \citenamefont {Mokrousov}, \citenamefont {Lee},\ and\
		\citenamefont {Ando}}]{OHE-OT-large}%
	\BibitemOpen
	\bibfield  {author} {\bibinfo {author} {\bibfnamefont {H.}~\bibnamefont
			{Hayashi}}, \bibinfo {author} {\bibfnamefont {D.}~\bibnamefont {Jo}},
		\bibinfo {author} {\bibfnamefont {D.}~\bibnamefont {Go}}, \bibinfo {author}
		{\bibfnamefont {T.}~\bibnamefont {Gao}}, \bibinfo {author} {\bibfnamefont
			{S.}~\bibnamefont {Haku}}, \bibinfo {author} {\bibfnamefont {Y.}~\bibnamefont
			{Mokrousov}}, \bibinfo {author} {\bibfnamefont {H.-W.}\ \bibnamefont {Lee}},\
		and\ \bibinfo {author} {\bibfnamefont {K.}~\bibnamefont {Ando}},\ }\href
	{https://doi.org/10.1038/s42005-023-01139-7} {\bibfield  {journal} {\bibinfo
			{journal} {Communications Physics}\ }\textbf {\bibinfo {volume} {6}},\
		\bibinfo {pages} {32} (\bibinfo {year} {2023})}\BibitemShut {NoStop}%
	\bibitem [{\citenamefont {Li}\ \emph {et~al.}(2023)\citenamefont {Li},
		\citenamefont {Liu}, \citenamefont {Li}, \citenamefont {Zhao}, \citenamefont
		{An},\ and\ \citenamefont {Ando}}]{L-S-OT-2023}%
	\BibitemOpen
	\bibfield  {author} {\bibinfo {author} {\bibfnamefont {T.}~\bibnamefont
			{Li}}, \bibinfo {author} {\bibfnamefont {L.}~\bibnamefont {Liu}}, \bibinfo
		{author} {\bibfnamefont {X.}~\bibnamefont {Li}}, \bibinfo {author}
		{\bibfnamefont {X.}~\bibnamefont {Zhao}}, \bibinfo {author} {\bibfnamefont
			{H.}~\bibnamefont {An}},\ and\ \bibinfo {author} {\bibfnamefont
			{K.}~\bibnamefont {Ando}},\ }\href
	{https://doi.org/10.1021/acs.nanolett.3c02104} {\bibfield  {journal}
		{\bibinfo  {journal} {Nano Letters}\ }\textbf {\bibinfo {volume} {23}},\
		\bibinfo {pages} {7174} (\bibinfo {year} {2023})},\ \bibinfo {note} {pMID:
		37466330},\ \Eprint
	{https://arxiv.org/abs/https://doi.org/10.1021/acs.nanolett.3c02104}
	{https://doi.org/10.1021/acs.nanolett.3c02104} \BibitemShut {NoStop}%
	\bibitem [{\citenamefont {et.al}(2022)}]{Exp-graphene-OHE-arXiv-2022}%
	\BibitemOpen
	\bibfield  {author} {\bibinfo {author} {\bibfnamefont {J.~S.-S.}\
			\bibnamefont {et.al}},\ }\href@noop {} {\bibfield  {journal} {\bibinfo
			{journal} {arXiv:2206.04565}\ } (\bibinfo {year} {2022})}\BibitemShut
	{NoStop}%
	\bibitem [{\citenamefont {Seifert}\ \emph {et~al.}(2023)\citenamefont
		{Seifert}, \citenamefont {Go}, \citenamefont {Hayashi}, \citenamefont
		{Rouzegar}, \citenamefont {Freimuth}, \citenamefont {Ando}, \citenamefont
		{Mokrousov},\ and\ \citenamefont {Kampfrath}}]{OAM-Exp-Tobias}%
	\BibitemOpen
	\bibfield  {author} {\bibinfo {author} {\bibfnamefont {T.~S.}\ \bibnamefont
			{Seifert}}, \bibinfo {author} {\bibfnamefont {D.}~\bibnamefont {Go}},
		\bibinfo {author} {\bibfnamefont {H.}~\bibnamefont {Hayashi}}, \bibinfo
		{author} {\bibfnamefont {R.}~\bibnamefont {Rouzegar}}, \bibinfo {author}
		{\bibfnamefont {F.}~\bibnamefont {Freimuth}}, \bibinfo {author}
		{\bibfnamefont {K.}~\bibnamefont {Ando}}, \bibinfo {author} {\bibfnamefont
			{Y.}~\bibnamefont {Mokrousov}},\ and\ \bibinfo {author} {\bibfnamefont
			{T.}~\bibnamefont {Kampfrath}},\ }\bibfield  {journal} {\bibinfo  {journal}
		{Nature Nanotechnology}\ }\href {https://doi.org/10.1038/s41565-023-01470-8}
	{10.1038/s41565-023-01470-8} (\bibinfo {year} {2023})\BibitemShut {NoStop}%
	\bibitem [{\citenamefont {\"Unzelmann}\ \emph {et~al.}(2020)\citenamefont
		{\"Unzelmann}, \citenamefont {Bentmann}, \citenamefont {Eck}, \citenamefont
		{Ki\ss{}linger}, \citenamefont {Geldiyev}, \citenamefont {Rieger},
		\citenamefont {Moser}, \citenamefont {Vidal}, \citenamefont {Ki\ss{}ner},
		\citenamefont {Hammer}, \citenamefont {Schneider}, \citenamefont {Fauster},
		\citenamefont {Sangiovanni}, \citenamefont {Di~Sante},\ and\ \citenamefont
		{Reinert}}]{OAM-PRL-2020-Reinert}%
	\BibitemOpen
	\bibfield  {author} {\bibinfo {author} {\bibfnamefont {M.}~\bibnamefont
			{\"Unzelmann}}, \bibinfo {author} {\bibfnamefont {H.}~\bibnamefont
			{Bentmann}}, \bibinfo {author} {\bibfnamefont {P.}~\bibnamefont {Eck}},
		\bibinfo {author} {\bibfnamefont {T.}~\bibnamefont {Ki\ss{}linger}}, \bibinfo
		{author} {\bibfnamefont {B.}~\bibnamefont {Geldiyev}}, \bibinfo {author}
		{\bibfnamefont {J.}~\bibnamefont {Rieger}}, \bibinfo {author} {\bibfnamefont
			{S.}~\bibnamefont {Moser}}, \bibinfo {author} {\bibfnamefont {R.~C.}\
			\bibnamefont {Vidal}}, \bibinfo {author} {\bibfnamefont {K.}~\bibnamefont
			{Ki\ss{}ner}}, \bibinfo {author} {\bibfnamefont {L.}~\bibnamefont {Hammer}},
		\bibinfo {author} {\bibfnamefont {M.~A.}\ \bibnamefont {Schneider}}, \bibinfo
		{author} {\bibfnamefont {T.}~\bibnamefont {Fauster}}, \bibinfo {author}
		{\bibfnamefont {G.}~\bibnamefont {Sangiovanni}}, \bibinfo {author}
		{\bibfnamefont {D.}~\bibnamefont {Di~Sante}},\ and\ \bibinfo {author}
		{\bibfnamefont {F.}~\bibnamefont {Reinert}},\ }\href
	{https://doi.org/10.1103/PhysRevLett.124.176401} {\bibfield  {journal}
		{\bibinfo  {journal} {Phys. Rev. Lett.}\ }\textbf {\bibinfo {volume} {124}},\
		\bibinfo {pages} {176401} (\bibinfo {year} {2020})}\BibitemShut {NoStop}%
	\bibitem [{\citenamefont {Salemi}\ \emph {et~al.}(2019)\citenamefont {Salemi},
		\citenamefont {Berritta}, \citenamefont {Nandy},\ and\ \citenamefont
		{Oppeneer}}]{OEE-NatComm-2019-Peter}%
	\BibitemOpen
	\bibfield  {author} {\bibinfo {author} {\bibfnamefont {L.}~\bibnamefont
			{Salemi}}, \bibinfo {author} {\bibfnamefont {M.}~\bibnamefont {Berritta}},
		\bibinfo {author} {\bibfnamefont {A.~K.}\ \bibnamefont {Nandy}},\ and\
		\bibinfo {author} {\bibfnamefont {P.~M.}\ \bibnamefont {Oppeneer}},\ }\href
	{https://doi.org/10.1038/s41467-019-13367-z} {\bibfield  {journal} {\bibinfo
			{journal} {Nature Communications}\ }\textbf {\bibinfo {volume} {10}},\
		\bibinfo {pages} {5381} (\bibinfo {year} {2019})}\BibitemShut {NoStop}%
	\bibitem [{\citenamefont {Tang}\ and\ \citenamefont {Bauer}(2024)}]{Tangping}%
	\BibitemOpen
	\bibfield  {author} {\bibinfo {author} {\bibfnamefont {P.}~\bibnamefont
			{Tang}}\ and\ \bibinfo {author} {\bibfnamefont {G.~E.~W.}\ \bibnamefont
			{Bauer}},\ }\href@noop {} {\bibfield  {journal} {\bibinfo  {journal}
			{arXiv:2401.17620}\ } (\bibinfo {year} {2024})}\BibitemShut {NoStop}%
	\bibitem [{\citenamefont {Wang}\ \emph
		{et~al.}(2023{\natexlab{b}})\citenamefont {Wang}, \citenamefont {Feng},
		\citenamefont {Yang}, \citenamefont {Zhang}, \citenamefont {Liu},
		\citenamefont {Xu}, \citenamefont {Jia}, \citenamefont {Wu}, \citenamefont
		{Yu}, \citenamefont {Xu},\ and\ \citenamefont
		{Jiang}}]{Inverse-OHE-weak-SOC}%
	\BibitemOpen
	\bibfield  {author} {\bibinfo {author} {\bibfnamefont {P.}~\bibnamefont
			{Wang}}, \bibinfo {author} {\bibfnamefont {Z.}~\bibnamefont {Feng}}, \bibinfo
		{author} {\bibfnamefont {Y.}~\bibnamefont {Yang}}, \bibinfo {author}
		{\bibfnamefont {D.}~\bibnamefont {Zhang}}, \bibinfo {author} {\bibfnamefont
			{Q.}~\bibnamefont {Liu}}, \bibinfo {author} {\bibfnamefont {Z.}~\bibnamefont
			{Xu}}, \bibinfo {author} {\bibfnamefont {Z.}~\bibnamefont {Jia}}, \bibinfo
		{author} {\bibfnamefont {Y.}~\bibnamefont {Wu}}, \bibinfo {author}
		{\bibfnamefont {G.}~\bibnamefont {Yu}}, \bibinfo {author} {\bibfnamefont
			{X.}~\bibnamefont {Xu}},\ and\ \bibinfo {author} {\bibfnamefont
			{Y.}~\bibnamefont {Jiang}},\ }\href
	{https://doi.org/10.1038/s41535-023-00559-6} {\bibfield  {journal} {\bibinfo
			{journal} {npj Quantum Materials}\ }\textbf {\bibinfo {volume} {8}},\
		\bibinfo {pages} {28} (\bibinfo {year} {2023}{\natexlab{b}})}\BibitemShut
	{NoStop}%
	\bibitem [{\citenamefont {Busch}\ \emph {et~al.}(2024)\citenamefont {Busch},
		\citenamefont {Ziolkowski}, \citenamefont {G\"obel}, \citenamefont {Mertig},\
		and\ \citenamefont {Henk}}]{PhysRevResearch.6.013208}%
	\BibitemOpen
	\bibfield  {author} {\bibinfo {author} {\bibfnamefont {O.}~\bibnamefont
			{Busch}}, \bibinfo {author} {\bibfnamefont {F.}~\bibnamefont {Ziolkowski}},
		\bibinfo {author} {\bibfnamefont {B.}~\bibnamefont {G\"obel}}, \bibinfo
		{author} {\bibfnamefont {I.}~\bibnamefont {Mertig}},\ and\ \bibinfo {author}
		{\bibfnamefont {J.}~\bibnamefont {Henk}},\ }\href
	{https://doi.org/10.1103/PhysRevResearch.6.013208} {\bibfield  {journal}
		{\bibinfo  {journal} {Phys. Rev. Res.}\ }\textbf {\bibinfo {volume} {6}},\
		\bibinfo {pages} {013208} (\bibinfo {year} {2024})}\BibitemShut {NoStop}%
	\bibitem [{\citenamefont {Ramaswamy}\ \emph {et~al.}(2018)\citenamefont
		{Ramaswamy}, \citenamefont {Lee}, \citenamefont {Cai},\ and\ \citenamefont
		{Yang}}]{Ramaswamy2018}%
	\BibitemOpen
	\bibfield  {author} {\bibinfo {author} {\bibfnamefont {R.}~\bibnamefont
			{Ramaswamy}}, \bibinfo {author} {\bibfnamefont {J.~M.}\ \bibnamefont {Lee}},
		\bibinfo {author} {\bibfnamefont {K.}~\bibnamefont {Cai}},\ and\ \bibinfo
		{author} {\bibfnamefont {H.}~\bibnamefont {Yang}},\ }\href
	{https://doi.org/10.1063/1.5041793} {\bibfield  {journal} {\bibinfo
			{journal} {Applied Physics Reviews}\ }\textbf {\bibinfo {volume} {5}},\
		\bibinfo {pages} {031107} (\bibinfo {year} {2018})},\ \Eprint
	{https://arxiv.org/abs/1808.06829} {arXiv:1808.06829} \BibitemShut {NoStop}%
	\bibitem [{\citenamefont {Manchon}\ \emph {et~al.}(2019)\citenamefont
		{Manchon}, \citenamefont {\ifmmode~\check{Z}\else \v{Z}\fi{}elezn\'y},
		\citenamefont {Miron}, \citenamefont {Jungwirth}, \citenamefont {Sinova},
		\citenamefont {Thiaville}, \citenamefont {Garello},\ and\ \citenamefont
		{Gambardella}}]{CI-SOT-RMP-2019-Manchon}%
	\BibitemOpen
	\bibfield  {author} {\bibinfo {author} {\bibfnamefont {A.}~\bibnamefont
			{Manchon}}, \bibinfo {author} {\bibfnamefont {J.}~\bibnamefont
			{\ifmmode~\check{Z}\else \v{Z}\fi{}elezn\'y}}, \bibinfo {author}
		{\bibfnamefont {I.~M.}\ \bibnamefont {Miron}}, \bibinfo {author}
		{\bibfnamefont {T.}~\bibnamefont {Jungwirth}}, \bibinfo {author}
		{\bibfnamefont {J.}~\bibnamefont {Sinova}}, \bibinfo {author} {\bibfnamefont
			{A.}~\bibnamefont {Thiaville}}, \bibinfo {author} {\bibfnamefont
			{K.}~\bibnamefont {Garello}},\ and\ \bibinfo {author} {\bibfnamefont
			{P.}~\bibnamefont {Gambardella}},\ }\href
	{https://doi.org/10.1103/RevModPhys.91.035004} {\bibfield  {journal}
		{\bibinfo  {journal} {Rev. Mod. Phys.}\ }\textbf {\bibinfo {volume} {91}},\
		\bibinfo {pages} {035004} (\bibinfo {year} {2019})}\BibitemShut {NoStop}%
	\bibitem [{\citenamefont {Fan}\ \emph {et~al.}(2017)\citenamefont {Fan},
		\citenamefont {Angizi},\ and\ \citenamefont {He}}]{fan2017memory}%
	\BibitemOpen
	\bibfield  {author} {\bibinfo {author} {\bibfnamefont {D.}~\bibnamefont
			{Fan}}, \bibinfo {author} {\bibfnamefont {S.}~\bibnamefont {Angizi}},\ and\
		\bibinfo {author} {\bibfnamefont {Z.}~\bibnamefont {He}},\ }in\ \href@noop {}
	{\emph {\bibinfo {booktitle} {2017 IEEE Computer Society Annual Symposium on
				VLSI (ISVLSI)}}}\ (\bibinfo {organization} {IEEE},\ \bibinfo {year} {2017})\
	pp.\ \bibinfo {pages} {683--688}\BibitemShut {NoStop}%
	\bibitem [{\citenamefont {Wang}\ \emph {et~al.}(2022)\citenamefont {Wang},
		\citenamefont {Sheng}, \citenamefont {Zheng}, \citenamefont {Ji},\ and\
		\citenamefont {Wang}}]{wang2022all}%
	\BibitemOpen
	\bibfield  {author} {\bibinfo {author} {\bibfnamefont {W.}~\bibnamefont
			{Wang}}, \bibinfo {author} {\bibfnamefont {Y.}~\bibnamefont {Sheng}},
		\bibinfo {author} {\bibfnamefont {Y.}~\bibnamefont {Zheng}}, \bibinfo
		{author} {\bibfnamefont {Y.}~\bibnamefont {Ji}},\ and\ \bibinfo {author}
		{\bibfnamefont {K.}~\bibnamefont {Wang}},\ }\href@noop {} {\bibfield
		{journal} {\bibinfo  {journal} {Advanced Electronic Materials}\ }\textbf
		{\bibinfo {volume} {8}},\ \bibinfo {pages} {2200412} (\bibinfo {year}
		{2022})}\BibitemShut {NoStop}%
	\bibitem [{\citenamefont {Marrows}\ \emph {et~al.}(2024)\citenamefont
		{Marrows}, \citenamefont {Barker}, \citenamefont {Moore},\ and\ \citenamefont
		{Moorsom}}]{marrows2024neuromorphic}%
	\BibitemOpen
	\bibfield  {author} {\bibinfo {author} {\bibfnamefont {C.~H.}\ \bibnamefont
			{Marrows}}, \bibinfo {author} {\bibfnamefont {J.}~\bibnamefont {Barker}},
		\bibinfo {author} {\bibfnamefont {T.~A.}\ \bibnamefont {Moore}},\ and\
		\bibinfo {author} {\bibfnamefont {T.}~\bibnamefont {Moorsom}},\ }\href@noop
	{} {\bibfield  {journal} {\bibinfo  {journal} {npj Spintronics}\ }\textbf
		{\bibinfo {volume} {2}},\ \bibinfo {pages} {12} (\bibinfo {year}
		{2024})}\BibitemShut {NoStop}%
	\bibitem [{\citenamefont {Grollier}\ \emph {et~al.}(2020)\citenamefont
		{Grollier}, \citenamefont {Querlioz}, \citenamefont {Camsari}, \citenamefont
		{Everschor-Sitte}, \citenamefont {Fukami},\ and\ \citenamefont
		{Stiles}}]{grollier2020neuromorphic}%
	\BibitemOpen
	\bibfield  {author} {\bibinfo {author} {\bibfnamefont {J.}~\bibnamefont
			{Grollier}}, \bibinfo {author} {\bibfnamefont {D.}~\bibnamefont {Querlioz}},
		\bibinfo {author} {\bibfnamefont {K.}~\bibnamefont {Camsari}}, \bibinfo
		{author} {\bibfnamefont {K.}~\bibnamefont {Everschor-Sitte}}, \bibinfo
		{author} {\bibfnamefont {S.}~\bibnamefont {Fukami}},\ and\ \bibinfo {author}
		{\bibfnamefont {M.~D.}\ \bibnamefont {Stiles}},\ }\href@noop {} {\bibfield
		{journal} {\bibinfo  {journal} {Nature electronics}\ }\textbf {\bibinfo
			{volume} {3}},\ \bibinfo {pages} {360} (\bibinfo {year} {2020})}\BibitemShut
	{NoStop}%
	\bibitem [{\citenamefont {Wang}\ and\ \citenamefont {Yang}(2022)}]{SOT-TM-Rev}%
	\BibitemOpen
	\bibfield  {author} {\bibinfo {author} {\bibfnamefont {Y.}~\bibnamefont
			{Wang}}\ and\ \bibinfo {author} {\bibfnamefont {H.}~\bibnamefont {Yang}},\
	}\href@noop {} {\bibfield  {journal} {\bibinfo  {journal} {Accounts of
				Materials Research}\ }\textbf {\bibinfo {volume} {3}},\ \bibinfo {pages}
		{1061} (\bibinfo {year} {2022})}\BibitemShut {NoStop}%
	\bibitem [{\citenamefont {Chang}\ \emph {et~al.}(2015)\citenamefont {Chang},
		\citenamefont {Markussen}, \citenamefont {Smidstrup}, \citenamefont
		{Stokbro},\ and\ \citenamefont {Nikoli{\'{c}}}}]{Chang2015}%
	\BibitemOpen
	\bibfield  {author} {\bibinfo {author} {\bibfnamefont {P.~H.}\ \bibnamefont
			{Chang}}, \bibinfo {author} {\bibfnamefont {T.}~\bibnamefont {Markussen}},
		\bibinfo {author} {\bibfnamefont {S.}~\bibnamefont {Smidstrup}}, \bibinfo
		{author} {\bibfnamefont {K.}~\bibnamefont {Stokbro}},\ and\ \bibinfo {author}
		{\bibfnamefont {B.~K.}\ \bibnamefont {Nikoli{\'{c}}}},\ }\href
	{https://doi.org/10.1103/PHYSREVB.92.201406/FIGURES/4/MEDIUM} {\bibfield
		{journal} {\bibinfo  {journal} {Physical Review B - Condensed Matter and
				Materials Physics}\ }\textbf {\bibinfo {volume} {92}},\ \bibinfo {pages}
		{201406} (\bibinfo {year} {2015})},\ \Eprint
	{https://arxiv.org/abs/1503.08046} {arXiv:1503.08046} \BibitemShut {NoStop}%
	\bibitem [{\citenamefont {Ghosh}\ and\ \citenamefont
		{Manchon}(2018)}]{PhysRevB.97.134402-Manchon}%
	\BibitemOpen
	\bibfield  {author} {\bibinfo {author} {\bibfnamefont {S.}~\bibnamefont
			{Ghosh}}\ and\ \bibinfo {author} {\bibfnamefont {A.}~\bibnamefont
			{Manchon}},\ }\href {https://doi.org/10.1103/PhysRevB.97.134402} {\bibfield
		{journal} {\bibinfo  {journal} {Phys. Rev. B}\ }\textbf {\bibinfo {volume}
			{97}},\ \bibinfo {pages} {134402} (\bibinfo {year} {2018})}\BibitemShut
	{NoStop}%
	\bibitem [{\citenamefont {Wang}\ \emph {et~al.}(2017)\citenamefont {Wang},
		\citenamefont {Zhu}, \citenamefont {Wu}, \citenamefont {Yang}, \citenamefont
		{Yu}, \citenamefont {Ramaswamy}, \citenamefont {Mishra}, \citenamefont {Shi},
		\citenamefont {Elyasi}, \citenamefont {Teo}, \citenamefont {Wu},\ and\
		\citenamefont {Yang}}]{TI-FM-Switch}%
	\BibitemOpen
	\bibfield  {author} {\bibinfo {author} {\bibfnamefont {Y.}~\bibnamefont
			{Wang}}, \bibinfo {author} {\bibfnamefont {D.}~\bibnamefont {Zhu}}, \bibinfo
		{author} {\bibfnamefont {Y.}~\bibnamefont {Wu}}, \bibinfo {author}
		{\bibfnamefont {Y.}~\bibnamefont {Yang}}, \bibinfo {author} {\bibfnamefont
			{J.}~\bibnamefont {Yu}}, \bibinfo {author} {\bibfnamefont {R.}~\bibnamefont
			{Ramaswamy}}, \bibinfo {author} {\bibfnamefont {R.}~\bibnamefont {Mishra}},
		\bibinfo {author} {\bibfnamefont {S.}~\bibnamefont {Shi}}, \bibinfo {author}
		{\bibfnamefont {M.}~\bibnamefont {Elyasi}}, \bibinfo {author} {\bibfnamefont
			{K.-L.}\ \bibnamefont {Teo}}, \bibinfo {author} {\bibfnamefont
			{Y.}~\bibnamefont {Wu}},\ and\ \bibinfo {author} {\bibfnamefont
			{H.}~\bibnamefont {Yang}},\ }\href
	{https://doi.org/10.1038/s41467-017-01583-4} {\bibfield  {journal} {\bibinfo
			{journal} {Nature Communications}\ }\textbf {\bibinfo {volume} {8}},\
		\bibinfo {pages} {1364} (\bibinfo {year} {2017})}\BibitemShut {NoStop}%
	\bibitem [{\citenamefont {Han}\ \emph {et~al.}(2017)\citenamefont {Han},
		\citenamefont {Richardella}, \citenamefont {Siddiqui}, \citenamefont
		{Finley}, \citenamefont {Samarth},\ and\ \citenamefont
		{Liu}}]{PhysRevLett.119.077702}%
	\BibitemOpen
	\bibfield  {author} {\bibinfo {author} {\bibfnamefont {J.}~\bibnamefont
			{Han}}, \bibinfo {author} {\bibfnamefont {A.}~\bibnamefont {Richardella}},
		\bibinfo {author} {\bibfnamefont {S.~A.}\ \bibnamefont {Siddiqui}}, \bibinfo
		{author} {\bibfnamefont {J.}~\bibnamefont {Finley}}, \bibinfo {author}
		{\bibfnamefont {N.}~\bibnamefont {Samarth}},\ and\ \bibinfo {author}
		{\bibfnamefont {L.}~\bibnamefont {Liu}},\ }\href
	{https://doi.org/10.1103/PhysRevLett.119.077702} {\bibfield  {journal}
		{\bibinfo  {journal} {Phys. Rev. Lett.}\ }\textbf {\bibinfo {volume} {119}},\
		\bibinfo {pages} {077702} (\bibinfo {year} {2017})}\BibitemShut {NoStop}%
	\bibitem [{\citenamefont {Khang}\ \emph {et~al.}(2018)\citenamefont {Khang},
		\citenamefont {Ueda},\ and\ \citenamefont {Hai}}]{TI-FM-Switch-NM}%
	\BibitemOpen
	\bibfield  {author} {\bibinfo {author} {\bibfnamefont {N.~H.~D.}\
			\bibnamefont {Khang}}, \bibinfo {author} {\bibfnamefont {Y.}~\bibnamefont
			{Ueda}},\ and\ \bibinfo {author} {\bibfnamefont {P.~N.}\ \bibnamefont
			{Hai}},\ }\href {https://doi.org/10.1038/s41563-018-0137-y} {\bibfield
		{journal} {\bibinfo  {journal} {Nature Materials}\ }\textbf {\bibinfo
			{volume} {17}},\ \bibinfo {pages} {808} (\bibinfo {year} {2018})}\BibitemShut
	{NoStop}%
	\bibitem [{\citenamefont {Wang}\ \emph
		{et~al.}(2023{\natexlab{c}})\citenamefont {Wang}, \citenamefont {Wu},
		\citenamefont {Zhang}, \citenamefont {Liu}, \citenamefont {Chen},
		\citenamefont {Pandey}, \citenamefont {Yin}, \citenamefont {Wei},
		\citenamefont {Lei}, \citenamefont {Shi} \emph {et~al.}}]{wang2023room}%
	\BibitemOpen
	\bibfield  {author} {\bibinfo {author} {\bibfnamefont {H.}~\bibnamefont
			{Wang}}, \bibinfo {author} {\bibfnamefont {H.}~\bibnamefont {Wu}}, \bibinfo
		{author} {\bibfnamefont {J.}~\bibnamefont {Zhang}}, \bibinfo {author}
		{\bibfnamefont {Y.}~\bibnamefont {Liu}}, \bibinfo {author} {\bibfnamefont
			{D.}~\bibnamefont {Chen}}, \bibinfo {author} {\bibfnamefont {C.}~\bibnamefont
			{Pandey}}, \bibinfo {author} {\bibfnamefont {J.}~\bibnamefont {Yin}},
		\bibinfo {author} {\bibfnamefont {D.}~\bibnamefont {Wei}}, \bibinfo {author}
		{\bibfnamefont {N.}~\bibnamefont {Lei}}, \bibinfo {author} {\bibfnamefont
			{S.}~\bibnamefont {Shi}}, \emph {et~al.},\ }\href@noop {} {\bibfield
		{journal} {\bibinfo  {journal} {Nature communications}\ }\textbf {\bibinfo
			{volume} {14}},\ \bibinfo {pages} {5173} (\bibinfo {year}
		{2023}{\natexlab{c}})}\BibitemShut {NoStop}%
	\bibitem [{\citenamefont {Cui}\ \emph {et~al.}(2023)\citenamefont {Cui},
		\citenamefont {Chen}, \citenamefont {Zhang}, \citenamefont {Fang},
		\citenamefont {Zeng}, \citenamefont {Zhang}, \citenamefont {Zhang},
		\citenamefont {He}, \citenamefont {Yu}, \citenamefont {Yan} \emph
		{et~al.}}]{cui2023low}%
	\BibitemOpen
	\bibfield  {author} {\bibinfo {author} {\bibfnamefont {B.}~\bibnamefont
			{Cui}}, \bibinfo {author} {\bibfnamefont {A.}~\bibnamefont {Chen}}, \bibinfo
		{author} {\bibfnamefont {X.}~\bibnamefont {Zhang}}, \bibinfo {author}
		{\bibfnamefont {B.}~\bibnamefont {Fang}}, \bibinfo {author} {\bibfnamefont
			{Z.}~\bibnamefont {Zeng}}, \bibinfo {author} {\bibfnamefont {P.}~\bibnamefont
			{Zhang}}, \bibinfo {author} {\bibfnamefont {J.}~\bibnamefont {Zhang}},
		\bibinfo {author} {\bibfnamefont {W.}~\bibnamefont {He}}, \bibinfo {author}
		{\bibfnamefont {G.}~\bibnamefont {Yu}}, \bibinfo {author} {\bibfnamefont
			{P.}~\bibnamefont {Yan}}, \emph {et~al.},\ }\href@noop {} {\bibfield
		{journal} {\bibinfo  {journal} {Advanced Materials}\ }\textbf {\bibinfo
			{volume} {35}},\ \bibinfo {pages} {2302350} (\bibinfo {year}
		{2023})}\BibitemShut {NoStop}%
	\bibitem [{\citenamefont {Kurebayashi}\ and\ \citenamefont
		{Nagaosa}(2019)}]{kurebayashi2019}%
	\BibitemOpen
	\bibfield  {author} {\bibinfo {author} {\bibfnamefont {D.}~\bibnamefont
			{Kurebayashi}}\ and\ \bibinfo {author} {\bibfnamefont {N.}~\bibnamefont
			{Nagaosa}},\ }\href@noop {} {\bibfield  {journal} {\bibinfo  {journal}
			{Physical Review B}\ }\textbf {\bibinfo {volume} {100}},\ \bibinfo {pages}
		{134407} (\bibinfo {year} {2019})}\BibitemShut {NoStop}%
	\bibitem [{\citenamefont {Jamali}\ \emph {et~al.}(2015)\citenamefont {Jamali},
		\citenamefont {Lee}, \citenamefont {Jeong}, \citenamefont {Mahfouzi},
		\citenamefont {Lv}, \citenamefont {Zhao}, \citenamefont {Nikoli{\'c}},
		\citenamefont {Mkhoyan}, \citenamefont {Samarth},\ and\ \citenamefont
		{Wang}}]{SOT-TI-bulk-NL}%
	\BibitemOpen
	\bibfield  {author} {\bibinfo {author} {\bibfnamefont {M.}~\bibnamefont
			{Jamali}}, \bibinfo {author} {\bibfnamefont {J.~S.}\ \bibnamefont {Lee}},
		\bibinfo {author} {\bibfnamefont {J.~S.}\ \bibnamefont {Jeong}}, \bibinfo
		{author} {\bibfnamefont {F.}~\bibnamefont {Mahfouzi}}, \bibinfo {author}
		{\bibfnamefont {Y.}~\bibnamefont {Lv}}, \bibinfo {author} {\bibfnamefont
			{Z.}~\bibnamefont {Zhao}}, \bibinfo {author} {\bibfnamefont {B.~K.}\
			\bibnamefont {Nikoli{\'c}}}, \bibinfo {author} {\bibfnamefont {K.~A.}\
			\bibnamefont {Mkhoyan}}, \bibinfo {author} {\bibfnamefont {N.}~\bibnamefont
			{Samarth}},\ and\ \bibinfo {author} {\bibfnamefont {J.-P.}\ \bibnamefont
			{Wang}},\ }\bibfield  {booktitle} {\emph {\bibinfo {booktitle} {Nano
				Letters}},\ }\href {https://doi.org/10.1021/acs.nanolett.5b03274} {\bibfield
		{journal} {\bibinfo  {journal} {Nano Letters}\ }\textbf {\bibinfo {volume}
			{15}},\ \bibinfo {pages} {7126} (\bibinfo {year} {2015})}\BibitemShut
	{NoStop}%
	\bibitem [{\citenamefont {Liu}\ \emph {et~al.}(2023)\citenamefont {Liu},
		\citenamefont {Cullen},\ and\ \citenamefont {Culcer}}]{Hong-PSHE-PRB}%
	\BibitemOpen
	\bibfield  {author} {\bibinfo {author} {\bibfnamefont {H.}~\bibnamefont
			{Liu}}, \bibinfo {author} {\bibfnamefont {J.~H.}\ \bibnamefont {Cullen}},\
		and\ \bibinfo {author} {\bibfnamefont {D.}~\bibnamefont {Culcer}},\ }\href
	{https://doi.org/10.1103/PhysRevB.108.195434} {\bibfield  {journal} {\bibinfo
			{journal} {Phys. Rev. B}\ }\textbf {\bibinfo {volume} {108}},\ \bibinfo
		{pages} {195434} (\bibinfo {year} {2023})}\BibitemShut {NoStop}%
	\bibitem [{\citenamefont {Ma}\ \emph {et~al.}(2024)\citenamefont {Ma},
		\citenamefont {Cullen}, \citenamefont {Monir}, \citenamefont {Rahman},\ and\
		\citenamefont {Culcer}}]{ma2024spin}%
	\BibitemOpen
	\bibfield  {author} {\bibinfo {author} {\bibfnamefont {H.}~\bibnamefont
			{Ma}}, \bibinfo {author} {\bibfnamefont {J.~H.}\ \bibnamefont {Cullen}},
		\bibinfo {author} {\bibfnamefont {S.}~\bibnamefont {Monir}}, \bibinfo
		{author} {\bibfnamefont {R.}~\bibnamefont {Rahman}},\ and\ \bibinfo {author}
		{\bibfnamefont {D.}~\bibnamefont {Culcer}},\ }\href@noop {} {\bibfield
		{journal} {\bibinfo  {journal} {npj Spintronics}\ }\textbf {\bibinfo {volume}
			{2}},\ \bibinfo {pages} {55} (\bibinfo {year} {2024})}\BibitemShut {NoStop}%
	\bibitem [{\citenamefont {Cullen}\ \emph {et~al.}(2023)\citenamefont {Cullen},
		\citenamefont {Atencia},\ and\ \citenamefont {Culcer}}]{James-SOT}%
	\BibitemOpen
	\bibfield  {author} {\bibinfo {author} {\bibfnamefont {J.~H.}\ \bibnamefont
			{Cullen}}, \bibinfo {author} {\bibfnamefont {R.~B.}\ \bibnamefont
			{Atencia}},\ and\ \bibinfo {author} {\bibfnamefont {D.}~\bibnamefont
			{Culcer}},\ }\href@noop {} {\bibfield  {journal} {\bibinfo  {journal}
			{Nanoscale}\ }\textbf {\bibinfo {volume} {15}},\ \bibinfo {pages} {8437}
		(\bibinfo {year} {2023})}\BibitemShut {NoStop}%
	\bibitem [{\citenamefont {Park}\ \emph {et~al.}(2012)\citenamefont {Park},
		\citenamefont {Han}, \citenamefont {Kim}, \citenamefont {Koh}, \citenamefont
		{Kim}, \citenamefont {Lee}, \citenamefont {Choi}, \citenamefont {Han},
		\citenamefont {Lee}, \citenamefont {Hur}, \citenamefont {Arita},
		\citenamefont {Shimada}, \citenamefont {Namatame},\ and\ \citenamefont
		{Taniguchi}}]{PhysRevLett.108.046805}%
	\BibitemOpen
	\bibfield  {author} {\bibinfo {author} {\bibfnamefont {S.~R.}\ \bibnamefont
			{Park}}, \bibinfo {author} {\bibfnamefont {J.}~\bibnamefont {Han}}, \bibinfo
		{author} {\bibfnamefont {C.}~\bibnamefont {Kim}}, \bibinfo {author}
		{\bibfnamefont {Y.~Y.}\ \bibnamefont {Koh}}, \bibinfo {author} {\bibfnamefont
			{C.}~\bibnamefont {Kim}}, \bibinfo {author} {\bibfnamefont {H.}~\bibnamefont
			{Lee}}, \bibinfo {author} {\bibfnamefont {H.~J.}\ \bibnamefont {Choi}},
		\bibinfo {author} {\bibfnamefont {J.~H.}\ \bibnamefont {Han}}, \bibinfo
		{author} {\bibfnamefont {K.~D.}\ \bibnamefont {Lee}}, \bibinfo {author}
		{\bibfnamefont {N.~J.}\ \bibnamefont {Hur}}, \bibinfo {author} {\bibfnamefont
			{M.}~\bibnamefont {Arita}}, \bibinfo {author} {\bibfnamefont
			{K.}~\bibnamefont {Shimada}}, \bibinfo {author} {\bibfnamefont
			{H.}~\bibnamefont {Namatame}},\ and\ \bibinfo {author} {\bibfnamefont
			{M.}~\bibnamefont {Taniguchi}},\ }\href
	{https://doi.org/10.1103/PhysRevLett.108.046805} {\bibfield  {journal}
		{\bibinfo  {journal} {Phys. Rev. Lett.}\ }\textbf {\bibinfo {volume} {108}},\
		\bibinfo {pages} {046805} (\bibinfo {year} {2012})}\BibitemShut {NoStop}%
	\bibitem [{\citenamefont {Osumi}\ \emph {et~al.}(2021)\citenamefont {Osumi},
		\citenamefont {Zhang},\ and\ \citenamefont {Murakami}}]{osumi2021kinetic}%
	\BibitemOpen
	\bibfield  {author} {\bibinfo {author} {\bibfnamefont {K.}~\bibnamefont
			{Osumi}}, \bibinfo {author} {\bibfnamefont {T.}~\bibnamefont {Zhang}},\ and\
		\bibinfo {author} {\bibfnamefont {S.}~\bibnamefont {Murakami}},\ }\href@noop
	{} {\bibfield  {journal} {\bibinfo  {journal} {Communications Physics}\
		}\textbf {\bibinfo {volume} {4}},\ \bibinfo {pages} {211} (\bibinfo {year}
		{2021})}\BibitemShut {NoStop}%
	\bibitem [{\citenamefont {Liu}\ \emph {et~al.}(2010)\citenamefont {Liu},
		\citenamefont {Qi}, \citenamefont {Zhang}, \citenamefont {Dai}, \citenamefont
		{Fang},\ and\ \citenamefont {Zhang}}]{TI-bulk-Parameters}%
	\BibitemOpen
	\bibfield  {author} {\bibinfo {author} {\bibfnamefont {C.-X.}\ \bibnamefont
			{Liu}}, \bibinfo {author} {\bibfnamefont {X.-L.}\ \bibnamefont {Qi}},
		\bibinfo {author} {\bibfnamefont {H.}~\bibnamefont {Zhang}}, \bibinfo
		{author} {\bibfnamefont {X.}~\bibnamefont {Dai}}, \bibinfo {author}
		{\bibfnamefont {Z.}~\bibnamefont {Fang}},\ and\ \bibinfo {author}
		{\bibfnamefont {S.-C.}\ \bibnamefont {Zhang}},\ }\href
	{https://doi.org/10.1103/PhysRevB.82.045122} {\bibfield  {journal} {\bibinfo
			{journal} {Phys. Rev. B}\ }\textbf {\bibinfo {volume} {82}},\ \bibinfo
		{pages} {045122} (\bibinfo {year} {2010})}\BibitemShut {NoStop}%
	\bibitem [{\citenamefont {Liu}\ \emph {et~al.}(2025)\citenamefont {Liu},
		\citenamefont {Cullen}, \citenamefont {Arovas},\ and\ \citenamefont
		{Culcer}}]{liu2024quantumcorrectionorbitalhall}%
	\BibitemOpen
	\bibfield  {author} {\bibinfo {author} {\bibfnamefont {H.}~\bibnamefont
			{Liu}}, \bibinfo {author} {\bibfnamefont {J.~H.}\ \bibnamefont {Cullen}},
		\bibinfo {author} {\bibfnamefont {D.~P.}\ \bibnamefont {Arovas}},\ and\
		\bibinfo {author} {\bibfnamefont {D.}~\bibnamefont {Culcer}},\ }\href
	{https://doi.org/10.1103/PhysRevLett.134.036304} {\bibfield  {journal}
		{\bibinfo  {journal} {Phys. Rev. Lett.}\ }\textbf {\bibinfo {volume} {134}},\
		\bibinfo {pages} {036304} (\bibinfo {year} {2025})}\BibitemShut {NoStop}%
	\bibitem [{\citenamefont {Go}\ \emph {et~al.}(2020)\citenamefont {Go},
		\citenamefont {Freimuth}, \citenamefont {Hanke}, \citenamefont {Xue},
		\citenamefont {Gomonay}, \citenamefont {Lee}, \citenamefont {Bl\"ugel},
		\citenamefont {Haney}, \citenamefont {Lee},\ and\ \citenamefont
		{Mokrousov}}]{CIAM-PRR-2020-Yuriy}%
	\BibitemOpen
	\bibfield  {author} {\bibinfo {author} {\bibfnamefont {D.}~\bibnamefont
			{Go}}, \bibinfo {author} {\bibfnamefont {F.}~\bibnamefont {Freimuth}},
		\bibinfo {author} {\bibfnamefont {J.-P.}\ \bibnamefont {Hanke}}, \bibinfo
		{author} {\bibfnamefont {F.}~\bibnamefont {Xue}}, \bibinfo {author}
		{\bibfnamefont {O.}~\bibnamefont {Gomonay}}, \bibinfo {author} {\bibfnamefont
			{K.-J.}\ \bibnamefont {Lee}}, \bibinfo {author} {\bibfnamefont
			{S.}~\bibnamefont {Bl\"ugel}}, \bibinfo {author} {\bibfnamefont {P.~M.}\
			\bibnamefont {Haney}}, \bibinfo {author} {\bibfnamefont {H.-W.}\ \bibnamefont
			{Lee}},\ and\ \bibinfo {author} {\bibfnamefont {Y.}~\bibnamefont
			{Mokrousov}},\ }\href {https://doi.org/10.1103/PhysRevResearch.2.033401}
	{\bibfield  {journal} {\bibinfo  {journal} {Phys. Rev. Res.}\ }\textbf
		{\bibinfo {volume} {2}},\ \bibinfo {pages} {033401} (\bibinfo {year}
		{2020})}\BibitemShut {NoStop}%
	\bibitem [{\citenamefont {Ojha}\ \emph {et~al.}(2023)\citenamefont {Ojha},
		\citenamefont {Chatterjee}, \citenamefont {Lin}, \citenamefont {Wu},
		\citenamefont {Tseng} \emph {et~al.}}]{ojha2023spin}%
	\BibitemOpen
	\bibfield  {author} {\bibinfo {author} {\bibfnamefont {D.~K.}\ \bibnamefont
			{Ojha}}, \bibinfo {author} {\bibfnamefont {R.}~\bibnamefont {Chatterjee}},
		\bibinfo {author} {\bibfnamefont {Y.-L.}\ \bibnamefont {Lin}}, \bibinfo
		{author} {\bibfnamefont {Y.-H.}\ \bibnamefont {Wu}}, \bibinfo {author}
		{\bibfnamefont {Y.-C.}\ \bibnamefont {Tseng}}, \emph {et~al.},\ }\href@noop
	{} {\bibfield  {journal} {\bibinfo  {journal} {Journal of Magnetism and
				Magnetic Materials}\ }\textbf {\bibinfo {volume} {572}},\ \bibinfo {pages}
		{170638} (\bibinfo {year} {2023})}\BibitemShut {NoStop}%
	\bibitem [{\citenamefont {Lyalin}\ and\ \citenamefont
		{Kawakami}(2024)}]{lyalin2024interface}%
	\BibitemOpen
	\bibfield  {author} {\bibinfo {author} {\bibfnamefont {I.}~\bibnamefont
			{Lyalin}}\ and\ \bibinfo {author} {\bibfnamefont {R.~K.}\ \bibnamefont
			{Kawakami}},\ }\href@noop {} {\bibfield  {journal} {\bibinfo  {journal}
			{Physical Review B}\ }\textbf {\bibinfo {volume} {110}},\ \bibinfo {pages}
		{104418} (\bibinfo {year} {2024})}\BibitemShut {NoStop}%
	\bibitem [{\citenamefont {Fukunaga}\ \emph {et~al.}(2023)\citenamefont
		{Fukunaga}, \citenamefont {Haku}, \citenamefont {Hayashi},\ and\
		\citenamefont {Ando}}]{PhysRevResearch.5.023054}%
	\BibitemOpen
	\bibfield  {author} {\bibinfo {author} {\bibfnamefont {R.}~\bibnamefont
			{Fukunaga}}, \bibinfo {author} {\bibfnamefont {S.}~\bibnamefont {Haku}},
		\bibinfo {author} {\bibfnamefont {H.}~\bibnamefont {Hayashi}},\ and\ \bibinfo
		{author} {\bibfnamefont {K.}~\bibnamefont {Ando}},\ }\href
	{https://doi.org/10.1103/PhysRevResearch.5.023054} {\bibfield  {journal}
		{\bibinfo  {journal} {Phys. Rev. Res.}\ }\textbf {\bibinfo {volume} {5}},\
		\bibinfo {pages} {023054} (\bibinfo {year} {2023})}\BibitemShut {NoStop}%
	\bibitem [{\citenamefont {Shi}\ \emph {et~al.}(2007)\citenamefont {Shi},
		\citenamefont {Vignale}, \citenamefont {Xiao},\ and\ \citenamefont
		{Niu}}]{Theroy-OM-PRL-2007-Qian}%
	\BibitemOpen
	\bibfield  {author} {\bibinfo {author} {\bibfnamefont {J.}~\bibnamefont
			{Shi}}, \bibinfo {author} {\bibfnamefont {G.}~\bibnamefont {Vignale}},
		\bibinfo {author} {\bibfnamefont {D.}~\bibnamefont {Xiao}},\ and\ \bibinfo
		{author} {\bibfnamefont {Q.}~\bibnamefont {Niu}},\ }\href
	{https://doi.org/10.1103/PhysRevLett.99.197202} {\bibfield  {journal}
		{\bibinfo  {journal} {Phys. Rev. Lett.}\ }\textbf {\bibinfo {volume} {99}},\
		\bibinfo {pages} {197202} (\bibinfo {year} {2007})}\BibitemShut {NoStop}%
	\bibitem [{\citenamefont {Khaetskii}(2006)}]{PhysRevLett.96.056602}%
	\BibitemOpen
	\bibfield  {author} {\bibinfo {author} {\bibfnamefont {A.}~\bibnamefont
			{Khaetskii}},\ }\href {https://doi.org/10.1103/PhysRevLett.96.056602}
	{\bibfield  {journal} {\bibinfo  {journal} {Phys. Rev. Lett.}\ }\textbf
		{\bibinfo {volume} {96}},\ \bibinfo {pages} {056602} (\bibinfo {year}
		{2006})}\BibitemShut {NoStop}%
	\bibitem [{\citenamefont {Gradhand}\ \emph {et~al.}(2010)\citenamefont
		{Gradhand}, \citenamefont {Fedorov}, \citenamefont {Zahn},\ and\
		\citenamefont {Mertig}}]{PhysRevLett.104.186403}%
	\BibitemOpen
	\bibfield  {author} {\bibinfo {author} {\bibfnamefont {M.}~\bibnamefont
			{Gradhand}}, \bibinfo {author} {\bibfnamefont {D.~V.}\ \bibnamefont
			{Fedorov}}, \bibinfo {author} {\bibfnamefont {P.}~\bibnamefont {Zahn}},\ and\
		\bibinfo {author} {\bibfnamefont {I.}~\bibnamefont {Mertig}},\ }\href
	{https://doi.org/10.1103/PhysRevLett.104.186403} {\bibfield  {journal}
		{\bibinfo  {journal} {Phys. Rev. Lett.}\ }\textbf {\bibinfo {volume} {104}},\
		\bibinfo {pages} {186403} (\bibinfo {year} {2010})}\BibitemShut {NoStop}%
	\bibitem [{\citenamefont {Ferreira}\ \emph {et~al.}(2014)\citenamefont
		{Ferreira}, \citenamefont {Rappoport}, \citenamefont {Cazalilla},\ and\
		\citenamefont {Castro~Neto}}]{PhysRevLett.112.066601}%
	\BibitemOpen
	\bibfield  {author} {\bibinfo {author} {\bibfnamefont {A.}~\bibnamefont
			{Ferreira}}, \bibinfo {author} {\bibfnamefont {T.~G.}\ \bibnamefont
			{Rappoport}}, \bibinfo {author} {\bibfnamefont {M.~A.}\ \bibnamefont
			{Cazalilla}},\ and\ \bibinfo {author} {\bibfnamefont {A.~H.}\ \bibnamefont
			{Castro~Neto}},\ }\href {https://doi.org/10.1103/PhysRevLett.112.066601}
	{\bibfield  {journal} {\bibinfo  {journal} {Phys. Rev. Lett.}\ }\textbf
		{\bibinfo {volume} {112}},\ \bibinfo {pages} {066601} (\bibinfo {year}
		{2014})}\BibitemShut {NoStop}%
	\bibitem [{\citenamefont {Cullen}\ and\ \citenamefont
		{Culcer}(2023{\natexlab{a}})}]{PhysRevB.108.245418}%
	\BibitemOpen
	\bibfield  {author} {\bibinfo {author} {\bibfnamefont {J.~H.}\ \bibnamefont
			{Cullen}}\ and\ \bibinfo {author} {\bibfnamefont {D.}~\bibnamefont
			{Culcer}},\ }\href {https://doi.org/10.1103/PhysRevB.108.245418} {\bibfield
		{journal} {\bibinfo  {journal} {Phys. Rev. B}\ }\textbf {\bibinfo {volume}
			{108}},\ \bibinfo {pages} {245418} (\bibinfo {year}
		{2023}{\natexlab{a}})}\BibitemShut {NoStop}%
	\bibitem [{\citenamefont {Culcer}\ \emph {et~al.}(2017)\citenamefont {Culcer},
		\citenamefont {Sekine},\ and\ \citenamefont
		{MacDonald}}]{Interband-Coherence-PRB-2017-Dimi}%
	\BibitemOpen
	\bibfield  {author} {\bibinfo {author} {\bibfnamefont {D.}~\bibnamefont
			{Culcer}}, \bibinfo {author} {\bibfnamefont {A.}~\bibnamefont {Sekine}},\
		and\ \bibinfo {author} {\bibfnamefont {A.~H.}\ \bibnamefont {MacDonald}},\
	}\href {https://doi.org/10.1103/PhysRevB.96.035106} {\bibfield  {journal}
		{\bibinfo  {journal} {Phys. Rev. B}\ }\textbf {\bibinfo {volume} {96}},\
		\bibinfo {pages} {035106} (\bibinfo {year} {2017})}\BibitemShut {NoStop}%
	\bibitem [{\citenamefont {Atencia}\ \emph {et~al.}(2022)\citenamefont
		{Atencia}, \citenamefont {Niu},\ and\ \citenamefont
		{Culcer}}]{JE-PRR-Rhonald-2022}%
	\BibitemOpen
	\bibfield  {author} {\bibinfo {author} {\bibfnamefont {R.~B.}\ \bibnamefont
			{Atencia}}, \bibinfo {author} {\bibfnamefont {Q.}~\bibnamefont {Niu}},\ and\
		\bibinfo {author} {\bibfnamefont {D.}~\bibnamefont {Culcer}},\ }\href
	{https://doi.org/10.1103/PhysRevResearch.4.013001} {\bibfield  {journal}
		{\bibinfo  {journal} {Phys. Rev. Res.}\ }\textbf {\bibinfo {volume} {4}},\
		\bibinfo {pages} {013001} (\bibinfo {year} {2022})}\BibitemShut {NoStop}%
	\bibitem [{\citenamefont {Tanaka}\ \emph {et~al.}(2008)\citenamefont {Tanaka},
		\citenamefont {Kontani}, \citenamefont {Naito}, \citenamefont {Naito},
		\citenamefont {Hirashima}, \citenamefont {Yamada},\ and\ \citenamefont
		{Inoue}}]{ISHE-IOHE-PRB-2008-Inoue}%
	\BibitemOpen
	\bibfield  {author} {\bibinfo {author} {\bibfnamefont {T.}~\bibnamefont
			{Tanaka}}, \bibinfo {author} {\bibfnamefont {H.}~\bibnamefont {Kontani}},
		\bibinfo {author} {\bibfnamefont {M.}~\bibnamefont {Naito}}, \bibinfo
		{author} {\bibfnamefont {T.}~\bibnamefont {Naito}}, \bibinfo {author}
		{\bibfnamefont {D.~S.}\ \bibnamefont {Hirashima}}, \bibinfo {author}
		{\bibfnamefont {K.}~\bibnamefont {Yamada}},\ and\ \bibinfo {author}
		{\bibfnamefont {J.}~\bibnamefont {Inoue}},\ }\href
	{https://doi.org/10.1103/PhysRevB.77.165117} {\bibfield  {journal} {\bibinfo
			{journal} {Phys. Rev. B}\ }\textbf {\bibinfo {volume} {77}},\ \bibinfo
		{pages} {165117} (\bibinfo {year} {2008})}\BibitemShut {NoStop}%
	\bibitem [{\citenamefont {Kontani}\ \emph
		{et~al.}(2009{\natexlab{a}})\citenamefont {Kontani}, \citenamefont {Tanaka},
		\citenamefont {Hirashima}, \citenamefont {Yamada},\ and\ \citenamefont
		{Inoue}}]{OHE-PRL-2009-Inoue}%
	\BibitemOpen
	\bibfield  {author} {\bibinfo {author} {\bibfnamefont {H.}~\bibnamefont
			{Kontani}}, \bibinfo {author} {\bibfnamefont {T.}~\bibnamefont {Tanaka}},
		\bibinfo {author} {\bibfnamefont {D.~S.}\ \bibnamefont {Hirashima}}, \bibinfo
		{author} {\bibfnamefont {K.}~\bibnamefont {Yamada}},\ and\ \bibinfo {author}
		{\bibfnamefont {J.}~\bibnamefont {Inoue}},\ }\href
	{https://doi.org/10.1103/PhysRevLett.102.016601} {\bibfield  {journal}
		{\bibinfo  {journal} {Phys. Rev. Lett.}\ }\textbf {\bibinfo {volume} {102}},\
		\bibinfo {pages} {016601} (\bibinfo {year} {2009}{\natexlab{a}})}\BibitemShut
	{NoStop}%
	\bibitem [{\citenamefont {Canonico}\ \emph {et~al.}(2020)\citenamefont
		{Canonico}, \citenamefont {Cysne}, \citenamefont {Rappoport},\ and\
		\citenamefont {Muniz}}]{canonico2020two}%
	\BibitemOpen
	\bibfield  {author} {\bibinfo {author} {\bibfnamefont {L.~M.}\ \bibnamefont
			{Canonico}}, \bibinfo {author} {\bibfnamefont {T.~P.}\ \bibnamefont {Cysne}},
		\bibinfo {author} {\bibfnamefont {T.~G.}\ \bibnamefont {Rappoport}},\ and\
		\bibinfo {author} {\bibfnamefont {R.}~\bibnamefont {Muniz}},\ }\href@noop {}
	{\bibfield  {journal} {\bibinfo  {journal} {Physical Review B}\ }\textbf
		{\bibinfo {volume} {101}},\ \bibinfo {pages} {075429} (\bibinfo {year}
		{2020})}\BibitemShut {NoStop}%
	\bibitem [{\citenamefont {Cullen}\ and\ \citenamefont
		{Culcer}(2023{\natexlab{b}})}]{cullen2023spin}%
	\BibitemOpen
	\bibfield  {author} {\bibinfo {author} {\bibfnamefont {J.~H.}\ \bibnamefont
			{Cullen}}\ and\ \bibinfo {author} {\bibfnamefont {D.}~\bibnamefont
			{Culcer}},\ }\href@noop {} {\bibfield  {journal} {\bibinfo  {journal}
			{Physical Review B}\ }\textbf {\bibinfo {volume} {108}},\ \bibinfo {pages}
		{245418} (\bibinfo {year} {2023}{\natexlab{b}})}\BibitemShut {NoStop}%
	\bibitem [{\citenamefont {Ado}\ \emph {et~al.}(2024)\citenamefont {Ado},
		\citenamefont {Titov}, \citenamefont {Duine},\ and\ \citenamefont
		{Brataas}}]{OEE-scalar-potential}%
	\BibitemOpen
	\bibfield  {author} {\bibinfo {author} {\bibfnamefont {I.~A.}\ \bibnamefont
			{Ado}}, \bibinfo {author} {\bibfnamefont {M.}~\bibnamefont {Titov}}, \bibinfo
		{author} {\bibfnamefont {R.~A.}\ \bibnamefont {Duine}},\ and\ \bibinfo
		{author} {\bibfnamefont {A.}~\bibnamefont {Brataas}},\ }\href
	{https://arxiv.org/abs/2407.00516} {\  (\bibinfo {year} {2024})},\ \Eprint
	{https://arxiv.org/abs/2407.00516} {arXiv:2407.00516 [cond-mat.mes-hall]}
	\BibitemShut {NoStop}%
	\bibitem [{\citenamefont {Wu}\ \emph {et~al.}(2019{\natexlab{a}})\citenamefont
		{Wu}, \citenamefont {Zhang}, \citenamefont {Deng}, \citenamefont {Lan},
		\citenamefont {Pan}, \citenamefont {Razavi}, \citenamefont {Che},
		\citenamefont {Huang}, \citenamefont {Dai}, \citenamefont {Wong},
		\citenamefont {Han},\ and\ \citenamefont {Wang}}]{PhysRevLett.123.207205}%
	\BibitemOpen
	\bibfield  {author} {\bibinfo {author} {\bibfnamefont {H.}~\bibnamefont
			{Wu}}, \bibinfo {author} {\bibfnamefont {P.}~\bibnamefont {Zhang}}, \bibinfo
		{author} {\bibfnamefont {P.}~\bibnamefont {Deng}}, \bibinfo {author}
		{\bibfnamefont {Q.}~\bibnamefont {Lan}}, \bibinfo {author} {\bibfnamefont
			{Q.}~\bibnamefont {Pan}}, \bibinfo {author} {\bibfnamefont {S.~A.}\
			\bibnamefont {Razavi}}, \bibinfo {author} {\bibfnamefont {X.}~\bibnamefont
			{Che}}, \bibinfo {author} {\bibfnamefont {L.}~\bibnamefont {Huang}}, \bibinfo
		{author} {\bibfnamefont {B.}~\bibnamefont {Dai}}, \bibinfo {author}
		{\bibfnamefont {K.}~\bibnamefont {Wong}}, \bibinfo {author} {\bibfnamefont
			{X.}~\bibnamefont {Han}},\ and\ \bibinfo {author} {\bibfnamefont {K.~L.}\
			\bibnamefont {Wang}},\ }\href
	{https://doi.org/10.1103/PhysRevLett.123.207205} {\bibfield  {journal}
		{\bibinfo  {journal} {Phys. Rev. Lett.}\ }\textbf {\bibinfo {volume} {123}},\
		\bibinfo {pages} {207205} (\bibinfo {year} {2019}{\natexlab{a}})}\BibitemShut
	{NoStop}%
	\bibitem [{\citenamefont {Dc}\ \emph {et~al.}(2018)\citenamefont {Dc},
		\citenamefont {Grassi}, \citenamefont {Chen}, \citenamefont {Jamali},
		\citenamefont {Reifsnyder~Hickey}, \citenamefont {Zhang}, \citenamefont
		{Zhao}, \citenamefont {Li}, \citenamefont {Quarterman}, \citenamefont {Lv}
		\emph {et~al.}}]{dc2018room}%
	\BibitemOpen
	\bibfield  {author} {\bibinfo {author} {\bibfnamefont {M.}~\bibnamefont
			{Dc}}, \bibinfo {author} {\bibfnamefont {R.}~\bibnamefont {Grassi}}, \bibinfo
		{author} {\bibfnamefont {J.-Y.}\ \bibnamefont {Chen}}, \bibinfo {author}
		{\bibfnamefont {M.}~\bibnamefont {Jamali}}, \bibinfo {author} {\bibfnamefont
			{D.}~\bibnamefont {Reifsnyder~Hickey}}, \bibinfo {author} {\bibfnamefont
			{D.}~\bibnamefont {Zhang}}, \bibinfo {author} {\bibfnamefont
			{Z.}~\bibnamefont {Zhao}}, \bibinfo {author} {\bibfnamefont {H.}~\bibnamefont
			{Li}}, \bibinfo {author} {\bibfnamefont {P.}~\bibnamefont {Quarterman}},
		\bibinfo {author} {\bibfnamefont {Y.}~\bibnamefont {Lv}}, \emph {et~al.},\
	}\href@noop {} {\bibfield  {journal} {\bibinfo  {journal} {Nature materials}\
		}\textbf {\bibinfo {volume} {17}},\ \bibinfo {pages} {800} (\bibinfo {year}
		{2018})}\BibitemShut {NoStop}%
	\bibitem [{\citenamefont {Wu}\ \emph {et~al.}(2021)\citenamefont {Wu},
		\citenamefont {Chen}, \citenamefont {Zhang}, \citenamefont {He},
		\citenamefont {Nance}, \citenamefont {Guo}, \citenamefont {Sasaki},
		\citenamefont {Shirokura}, \citenamefont {Hai}, \citenamefont {Fang} \emph
		{et~al.}}]{wu2021magnetic}%
	\BibitemOpen
	\bibfield  {author} {\bibinfo {author} {\bibfnamefont {H.}~\bibnamefont
			{Wu}}, \bibinfo {author} {\bibfnamefont {A.}~\bibnamefont {Chen}}, \bibinfo
		{author} {\bibfnamefont {P.}~\bibnamefont {Zhang}}, \bibinfo {author}
		{\bibfnamefont {H.}~\bibnamefont {He}}, \bibinfo {author} {\bibfnamefont
			{J.}~\bibnamefont {Nance}}, \bibinfo {author} {\bibfnamefont
			{C.}~\bibnamefont {Guo}}, \bibinfo {author} {\bibfnamefont {J.}~\bibnamefont
			{Sasaki}}, \bibinfo {author} {\bibfnamefont {T.}~\bibnamefont {Shirokura}},
		\bibinfo {author} {\bibfnamefont {P.~N.}\ \bibnamefont {Hai}}, \bibinfo
		{author} {\bibfnamefont {B.}~\bibnamefont {Fang}}, \emph {et~al.},\
	}\href@noop {} {\bibfield  {journal} {\bibinfo  {journal} {Nature
				communications}\ }\textbf {\bibinfo {volume} {12}},\ \bibinfo {pages} {6251}
		(\bibinfo {year} {2021})}\BibitemShut {NoStop}%
	\bibitem [{\citenamefont {Wu}\ \emph {et~al.}(2019{\natexlab{b}})\citenamefont
		{Wu}, \citenamefont {Zhang}, \citenamefont {Deng}, \citenamefont {Lan},
		\citenamefont {Pan}, \citenamefont {Razavi}, \citenamefont {Che},
		\citenamefont {Huang}, \citenamefont {Dai}, \citenamefont {Wong} \emph
		{et~al.}}]{wu2019room}%
	\BibitemOpen
	\bibfield  {author} {\bibinfo {author} {\bibfnamefont {H.}~\bibnamefont
			{Wu}}, \bibinfo {author} {\bibfnamefont {P.}~\bibnamefont {Zhang}}, \bibinfo
		{author} {\bibfnamefont {P.}~\bibnamefont {Deng}}, \bibinfo {author}
		{\bibfnamefont {Q.}~\bibnamefont {Lan}}, \bibinfo {author} {\bibfnamefont
			{Q.}~\bibnamefont {Pan}}, \bibinfo {author} {\bibfnamefont {S.~A.}\
			\bibnamefont {Razavi}}, \bibinfo {author} {\bibfnamefont {X.}~\bibnamefont
			{Che}}, \bibinfo {author} {\bibfnamefont {L.}~\bibnamefont {Huang}}, \bibinfo
		{author} {\bibfnamefont {B.}~\bibnamefont {Dai}}, \bibinfo {author}
		{\bibfnamefont {K.}~\bibnamefont {Wong}}, \emph {et~al.},\ }\href@noop {}
	{\bibfield  {journal} {\bibinfo  {journal} {Physical review letters}\
		}\textbf {\bibinfo {volume} {123}},\ \bibinfo {pages} {207205} (\bibinfo
		{year} {2019}{\natexlab{b}})}\BibitemShut {NoStop}%
	\bibitem [{\citenamefont {Kontani}\ \emph
		{et~al.}(2009{\natexlab{b}})\citenamefont {Kontani}, \citenamefont {Tanaka},
		\citenamefont {Hirashima}, \citenamefont {Yamada},\ and\ \citenamefont
		{Inoue}}]{PhysRevLett.102.016601}%
	\BibitemOpen
	\bibfield  {author} {\bibinfo {author} {\bibfnamefont {H.}~\bibnamefont
			{Kontani}}, \bibinfo {author} {\bibfnamefont {T.}~\bibnamefont {Tanaka}},
		\bibinfo {author} {\bibfnamefont {D.~S.}\ \bibnamefont {Hirashima}}, \bibinfo
		{author} {\bibfnamefont {K.}~\bibnamefont {Yamada}},\ and\ \bibinfo {author}
		{\bibfnamefont {J.}~\bibnamefont {Inoue}},\ }\href
	{https://doi.org/10.1103/PhysRevLett.102.016601} {\bibfield  {journal}
		{\bibinfo  {journal} {Phys. Rev. Lett.}\ }\textbf {\bibinfo {volume} {102}},\
		\bibinfo {pages} {016601} (\bibinfo {year} {2009}{\natexlab{b}})}\BibitemShut
	{NoStop}%
	\bibitem [{\citenamefont {Go}\ \emph {et~al.}(2018{\natexlab{b}})\citenamefont
		{Go}, \citenamefont {Jo}, \citenamefont {Kim},\ and\ \citenamefont
		{Lee}}]{PhysRevLett.121.086602}%
	\BibitemOpen
	\bibfield  {author} {\bibinfo {author} {\bibfnamefont {D.}~\bibnamefont
			{Go}}, \bibinfo {author} {\bibfnamefont {D.}~\bibnamefont {Jo}}, \bibinfo
		{author} {\bibfnamefont {C.}~\bibnamefont {Kim}},\ and\ \bibinfo {author}
		{\bibfnamefont {H.-W.}\ \bibnamefont {Lee}},\ }\href
	{https://doi.org/10.1103/PhysRevLett.121.086602} {\bibfield  {journal}
		{\bibinfo  {journal} {Phys. Rev. Lett.}\ }\textbf {\bibinfo {volume} {121}},\
		\bibinfo {pages} {086602} (\bibinfo {year} {2018}{\natexlab{b}})}\BibitemShut
	{NoStop}%
	\bibitem [{\citenamefont {Park}\ \emph {et~al.}(2011)\citenamefont {Park},
		\citenamefont {Kim}, \citenamefont {Yu}, \citenamefont {Han},\ and\
		\citenamefont {Kim}}]{OAM-Rashba-PRL-2011-Changyoung}%
	\BibitemOpen
	\bibfield  {author} {\bibinfo {author} {\bibfnamefont {S.~R.}\ \bibnamefont
			{Park}}, \bibinfo {author} {\bibfnamefont {C.~H.}\ \bibnamefont {Kim}},
		\bibinfo {author} {\bibfnamefont {J.}~\bibnamefont {Yu}}, \bibinfo {author}
		{\bibfnamefont {J.~H.}\ \bibnamefont {Han}},\ and\ \bibinfo {author}
		{\bibfnamefont {C.}~\bibnamefont {Kim}},\ }\href
	{https://doi.org/10.1103/PhysRevLett.107.156803} {\bibfield  {journal}
		{\bibinfo  {journal} {Phys. Rev. Lett.}\ }\textbf {\bibinfo {volume} {107}},\
		\bibinfo {pages} {156803} (\bibinfo {year} {2011})}\BibitemShut {NoStop}%
	\bibitem [{\citenamefont {Tatara}(2018)}]{Tatara-PRB-2018}%
	\BibitemOpen
	\bibfield  {author} {\bibinfo {author} {\bibfnamefont {G.}~\bibnamefont
			{Tatara}},\ }\href {https://doi.org/10.1103/PhysRevB.98.174422} {\bibfield
		{journal} {\bibinfo  {journal} {Phys. Rev. B}\ }\textbf {\bibinfo {volume}
			{98}},\ \bibinfo {pages} {174422} (\bibinfo {year} {2018})}\BibitemShut
	{NoStop}%
	\bibitem [{\citenamefont {Shitade}\ and\ \citenamefont
		{Tatara}(2022)}]{Tatara-PRB-Letter}%
	\BibitemOpen
	\bibfield  {author} {\bibinfo {author} {\bibfnamefont {A.}~\bibnamefont
			{Shitade}}\ and\ \bibinfo {author} {\bibfnamefont {G.}~\bibnamefont
			{Tatara}},\ }\href {https://doi.org/10.1103/PhysRevB.105.L201202} {\bibfield
		{journal} {\bibinfo  {journal} {Phys. Rev. B}\ }\textbf {\bibinfo {volume}
			{105}},\ \bibinfo {pages} {L201202} (\bibinfo {year} {2022})}\BibitemShut
	{NoStop}%
	\bibitem [{\citenamefont {Rhonald
			Burgos~Atencia}(2024)}]{Rhonald-Conservation-OMM}%
	\BibitemOpen
	\bibfield  {author} {\bibinfo {author} {\bibfnamefont {D.~C.}\ \bibnamefont
			{Rhonald Burgos~Atencia}, \bibfnamefont {Daniel P.~Arovas}},\ }\href@noop {}
	{\bibfield  {journal} {\bibinfo  {journal} {arXiv:2311.12108}\ } (\bibinfo
		{year} {2024})}\BibitemShut {NoStop}%
\end{thebibliography}
\end{document}